\documentclass[10pt,journal,compsoc]{IEEEtran}
\pdfoutput=1

\usepackage{amsmath}
\usepackage{amssymb}
\usepackage{amsthm}
\usepackage[pdftex]{graphicx}
\graphicspath{ {imgs/} }
\graphicspath{ {perfplot/} }
\usepackage{multicol}
\usepackage{blindtext} 
\usepackage{listings} 
\usepackage{subcaption}
\usepackage{siunitx}
\usepackage{float}
\restylefloat{table}

\usepackage{hyperref}
\hypersetup{
    colorlinks=true,
    linkcolor=blue,
    filecolor=magenta,      
    urlcolor=cyan,
}
\usepackage[table,xcdraw]{xcolor}
\usepackage{color}
\usepackage{textgreek}
\usepackage{pgfplots}
\usepackage[square,sort,comma,numbers]{natbib}
\usepackage{mathtools}

\usepackage{tikz}
\usetikzlibrary{calc}
\usetikzlibrary{arrows, decorations.markings}
\usetikzlibrary{shapes.geometric, fit, shapes, backgrounds, positioning, decorations.markings}
\tikzstyle{process} = [circle,text centered, draw=black, fill=red!20]
\tikzstyle{arrow} = [thick,->,>=stealth]
\tikzstyle{number} = [circle,text centered, draw=white, fill=white]

\tikzset{
  treenode/.style = {align=center, inner sep=0pt, text centered,
    font=\sffamily},
  root_style/.style = {treenode, circle, white, font=\sffamily\bfseries, draw=black,
    fill=black, text width=1.5em},
  child_style/.style = {treenode, circle, red,  draw=red, 
    text width=1.5em, very thick},
  arn_x/.style = {treenode, rectangle, draw=black,
    minimum width=0.5em, minimum height=0.5em},
  net connect/.style = {line width=1pt, draw=blue!50!cyan!25!black},
}

\usepackage{algorithm}
\usepackage{algorithmic}

\begin{document}

\title{Accurate runtime selection of optimal MPI collective algorithms using analytical performance modelling}

\author{Emin~Nuriyev,
        Alexey~Lastovetsky 
\thanks{E. Nuriyev and A. Lastovetsky are with the School of Computer Science, University College Dublin, Belfield, Dublin 4, Ireland. \protect\\
E-mail: emin.nuriyev@ucdconnect.ie, alexey.lastovetsky@ucd.ie}
}



\IEEEtitleabstractindextext{%

\begin{abstract} 
The performance of collective operations has been a critical issue since the advent of MPI. Many algorithms have been proposed for each MPI collective operation but none of them proved optimal in all situations. Different algorithms demonstrate superior performance depending on the platform, the message size, the number of processes, etc. MPI implementations perform the selection of the collective algorithm empirically, executing a simple runtime decision function. While efficient, this approach does not guarantee the optimal selection. As a more accurate but equally efficient alternative, the use of analytical performance models of collective algorithms for the selection process was proposed and studied. Unfortunately, the previous attempts in this direction have not been successful.

We revisit the analytical model-based approach and propose two innovations that significantly improve the selective accuracy of analytical models: (1) We derive analytical models from the code implementing the algorithms rather than from their high-level mathematical definitions. This results in more detailed models. (2) We estimate model parameters separately for each collective algorithm and include the execution of this algorithm in the corresponding communication experiment.

We experimentally demonstrate the accuracy and efficiency of our approach using Open MPI broadcast and gather algorithms and a Grid'5000 cluster.

\end{abstract}

}

\maketitle

\IEEEdisplaynontitleabstractindextext
\IEEEpeerreviewmaketitle

%
%
\ifCLASSOPTIONcompsoc
\IEEEraisesectionheading{\section{Introduction}\label{sec:intro}}
\else
\section{Introduction}
\label{sec:intro}
\fi
The message passing interface (MPI) \cite{mpi2012mpi} is the de-facto standard, which provides a reliable and portable environment for developing high-performance parallel applications on different platforms. Since the release of the first version of MPI, it provides a flexible communication layer including a mechanism for collective operations. MPI collective operations are classified into the following categories \cite{mpi2012mpi}: 1) \textbf{All-To-All} \textit{(MPI\_Allgather, MPI\_Alltoall, MPI\_Allreduce, MPI\_Barrier)}; 2) \textbf{All-To-One} \textit{(MPI\_Gather, MPI\_Reduce)}; 3) \textbf{One-To-All} \textit{(MPI\_Bcast, MPI\_Scatter)}; 4) \textbf{Other} \textit{(MPI\_Scan, MPI\_Exscan)}.

Rabenseifner \cite{rabenseifner1999automatic} shows that collective operations consume more than eighty percent of the total execution time of a typical MPI application. Therefore, a significant amount of research has been invested into optimisation of MPI collectives. Those researches have resulted in a large number of algorithms, each of which comes up optimal for specific message sizes, platforms, numbers of processes, and so forth. Mainstream MPI libraries provide multiple collective algorithms for each collective routine. For example, MPICH \cite{mpichlib} employs three broadcast algorithms to implement \textit{MPI\_Bcast}. In Open MPI library \cite{openmpilib}, the broadcast routine is built up with six different algorithms. However, none of the algorithms is optimal in all situations. Thus, there is a problem of selection of the optimal algorithm for each call of a collective routine, which normally depends on the platform, the number of processes, the message size and so forth.\par

There are two ways how this selection can be made in the MPI program. The first one, MPI\_T interface \cite{mpi2012mpi}, is provided by the MPI standard and allows the MPI programmer to select the collective algorithm explicitly from the list of available algorithms for each collective call at run-time. It does not solve the problem of optimal selection delegating its solution to the programmer.
The second one is transparent to the MPI programmer and provided by MPI implementations. It uses a simple \textit{decision function} in each collective routine, which is used to select the algorithm at runtime. The decision function is empirically derived from extensive testing on the dedicated system. For example, for each collective operation, both MPICH and Open MPI use a simple decision routine selecting the algorithm based on the message size and number of processes \cite{thakur2005optimization, gabriel2004open, fagg2006flexible}. The main advantage of this solution is its efficiency. The algorithm selection is very fast and does not affect the performance of the program. The main disadvantage of the existing decision functions is that they do not  guarantee the optimal selection in all situations.  \par

As an alternative approach, the use of analytical performance models of collective algorithms for the selection process has been proposed and studied. In the case of success, the analytical performance modelling approach, being as efficient as the existing decision functions approach, would guarantee the optimal selection in all situations. This approach was first proposed in \cite{pjevsivac2007performance}. In this work, several point-to-point communication models, such as Hockney \cite{hockney1994communication}, LogP \cite{culler1993logp}, LogGP \cite{alexandrov1995loggp}, PLogP \cite{kielmann2000fast}, are used to build analytical performance models of collective algorithms.  The analytical performance models are then used in decision functions for selection of the optimal algorithm. Unfortunately, the analytical performance models proposed in this work could not reach the level of accuracy sufficient for selection of the optimal algorithm. \par  


In this paper, we revisit the model-based approach and propose a number of innovations that significantly improve the selective accuracy of analytical models to the extent that allows them to be used for accurate selection of optimal collective algorithms. Our analytical modelling approach is based on the following innovations:

\begin{enumerate}
\item While previous attempts to build analytical performance models of collective algorithms only take into account their high-level mathematical definition, we derive our analytical models from the code implementing the algorithms. This results in much more detailed  models, which are able to correctly  compare  the performance of different algorithms implementing the same collective operation. 
\item  We propose to estimate the model parameters separately for each collective algorithm and carefully design the communication experiments for their estimation. 
More specifically, we design a specific communication experiment for each collective algorithm, so that the algorithm itself would be involved in the execution of the experiment. Moreover, the execution time of this experiment must be dominated by the execution time of this collective algorithm. Then, we conduct a number of experiments on the target platform for a range of numbers of processors and message sizes and accurately mesaure their execution times. From these experiments, we derive a sufficiently large number of equations with the model parameters as unknowns. Finally, we use a solver to find the values of the model parameters.
\end{enumerate}

We applied our approach to collective algorithms implemented in Open MPI. As a result, we managed to build a detailed analytical performance model for each collective algorithm and successfully use the models for selection of the optimal one. The accuracy of our solution has been validated on the Grid'5000 platform. \par

The main contributions of this paper can be summarized as follows:
\begin{itemize}
\item We propose and implement a new analytical performance modelling approach for MPI collective algorithms, which derives the models from the code implementing the algorithms.
\item  We propose and implement a novel approach to estimation of the parameters of analytical performance models of MPI collective algorithms, which estimates the parameters separately for each algorithm and includes the modelled collective algorithm in the communication experiment, which is used to estimate the model parameters. 
\item We experimentally validate the proposed approach to selection of optimal collective algorithms on the Grid'5000 platform \cite{grid5000}.  

\end{itemize} 

The rest of the paper is structured as follows. Section \ref{sec:relatedwork} reviews the existing approaches to performance modelling and algorithm selection problems. Section \ref{sec:collalgorithms} introduces MPI collective algorithms implemented in Open MPI. Section \ref{sec:implementationdrivenmodel} describes our approach to construction of analytical  performance models of MPI collective algorithms by deriving them from the MPI implementation. Section \ref{sec:designofcommexperiments} presents our method to measure analytical model parameters. Section \ref{sec:experimentalresults} presents experimental validation of the proposed  approach. Section \ref{sec:conclusions} concludes the paper with a discussion of the results and an outline of the future work.

\section{Related Work}
\label{sec:relatedwork}

In order to select the optimal algorithm for a given collective operation, we have to be able to accurately compare the performance of the available algorithms. Analytical performance models are one of the efficient ways to express and compare the performance of collective algorithms. In this section, we overview the state-of-the-art  in analytical performance modelling and measurement of model parameters. 

\subsection{Analytical performance models of MPI collective algorithms}
\label{subsec:analyticalperformancemodel}

All analytical models of collective algorithms use point-to-point communication models as building blocks. The most popular point-to-point communication models used in collective models are the Hockney model \cite{hockney1994communication}, LogP \cite{culler1993logp}, LogGP \cite{alexandrov1995loggp}, and PLogP \cite{kielmann2000fast}. In our work, we use the Hockney model, which estimates the time $T(m)$ of sending a message of size $m$ between two nodes as $ T(m) = \alpha + \beta \cdot m $, where $\alpha$ and  $\beta$ are the message latency and the reciprocal bandwidth respectively.

Thakur et al. \cite{thakur2005optimization} propose analytical performance models of several collective algorithms for \textit{MPI\_Allgather, MPI\_Bcast, MPI\_Alltoall, MPI\_Reduce\_scatter, MPI\_Reduce}, and \textit{MPI\_Allreduce} routines using the Hockney model. The parameters of the models, $\alpha$ and $\beta$, are assumed to be the same for all algorithms, message sizes and numbers of processes. The authors find their models not accurate enough for the task of selection of optimal collective algorithms. They conclude that in order to improve the accuracy of their analytical models, we have to assume that $\alpha$ and $\beta$ depend on the message size and the number of processes. They do not propose models improved this way though. In our work, we stick to the assumption of independence of model parameters on the message size and the number of processes. Instead, we improve the accuracy of our models by deriving them from the implementation of the modelled algorithms. In addition, we assume that $\alpha$ and $\beta$ may depend on the algorithm. Thus, our approach to improving the accuracy of models of collective algorithms is to make them more algorithm and implementation specific. 

Chan et al. \cite{chan2004optimizing} build analytical performance models of \textit{Minimum-spanning tree} algorithms and \textit{Bucket} algorithms for MPI\_Bcast, MPI\_Redcue, MPI\_Scatter, MPI\_Gather, MPI\_Allgather, MPI\_Reduce\_scatter, MPI\_Allreduce collectives and later extend this work for multidimensional mesh architecture in \cite{chan2007collective}. The proposed models are built using high-level theoretical descriptions of the algorithms. Therefore, the authors conclude that while the models can be used for analysis of theoretical complexity of the algorithms, they are not accurate enough for the task of estimation and comparison of their practical performance.

An analytical performance model of a new \textit{reduction} algorithm is proposed for a non-power-of-two number of processes by Rabenseifner et al. \cite{rabenseifner2004more}. The model uses a traditional high-level mathematical description of the algorithm. The aim of the model is to understand and express the complexity of the algorithm. Like in all previous models, its level of abstraction is too high to reach the accuracy required for comparison of the practical performance of the proposed reduction algorithm with its counterparts.

A general analytical performance model for \textit{tree-based} broadcast algorithms with message segmentation has been proposed by Patarasuk et al. \cite{patarasuk2006pipelined}. Unlike traditional models, this model introduces a new parameter, \textit{Maximum nodal degree} of the tree. The purpose of this  model is restricted to theoretical comparison of different tree-based broadcast algorithms. Accurate prediction of the execution time of the broadcast algorithms and methods for measurement of the model parameters, including the maximal nodal degree of the tree, are out of the scope of their work.   

Pjevsivac-Grbovic et al. \cite{pjevsivac2007performance} study selection of optimal collective algorithms using analytical performance models for \textit{barrier}, \textit{broadcast}, \textit{reduce} and \textit{alltoall} collective operations. Analytical performance models are built using the Hockney, LogP/LogGP, and PLogP point-to-point communication models. Additionally, the splitted-binary broadcast algorithm has been designed and analysed with different performance models in this work. The models are built up with the traditional approach using high-level mathematical definitions of the collective algorithms. In order to predict the cost of a collective algorithm by analytical formula, model parameters are measured using point-to-point communication experiments. After experimental validation of their modelling approach, the authors conclude that the proposed models are not accurate enough for selection of optimal algorithms. 

Lastovetsky et al. \cite{lastovetsky2006accurate} propose a point-to-point communication model for heterogeneous clusters. The model assumes that time to transmit a message between two nodes in a heterogeneous cluster is composed of the network transmission delay, source and destination processing delays. The analytical performance model of the binomial broadcast algorithm is built up using this model taking into account the impact of message passing protocols. While the predicted execution time of the binomial broadcast algorithm was close to the experimentally measured time, its use for comparison of practical performance of broadcast algorithms has never been studied.

\subsection{Measurement of model parameters}

One of the uses of analytical communication performance models is for theoretical analysis of the complexity of collective algorithms. In such purely theoretical studies, the authors do not pay much attention to methods of measurement of model parameters. However, if a model is intended for accurate prediction of the execution time of the communication algorithm on each particular platform, a well-defined experimental measurement method of the model parameters will be as important as the theoretical formulation of the model. Different measurement methods may give significantly different values of the model parameters and therefore either degrade or improve the model's prediction accuracy.
  
In general, a typical measurement method consists of a well-defined set of communication experiments, each of which is used to obtain an equation with model parameters as unknowns on one side of the equation and the measured execution time of the experiment on the other side. The full system of such equations is then solved to find the values of the model parameters for each particular platform. Existing measurement methods predominantly consist of \textit{point-to-point} communication experiments, which are used to obtain a system of \textit{linear} equations. In this subsection, we overview some notable works in this area.

Hockney \cite{hockney1994communication} presents a measurement method to find the $\alpha$ and $\beta$ parameters of the Hockney model. The set of communication experiments consists of point-to-point round-trips. The sender sends a message of size $m$ to the receiver, which immediately returns the message to the sender upon its receipt. The time $RTT(m)$ of this experiment is measured on the sender side and estimated as $RTT(m) = 2 \cdot (\alpha + m\cdot \beta)$. These round-trip communication experiments for a wide range of message size $m$ produce a system of linear equations with $\alpha$ and $\beta$ as unknowns. To find $\alpha$ and $\beta$ from this system, the linear least-squares regression is used.

Culler et al. \cite{culler1996logp} propose a method of measurement of parameters of the LogP model, namely,  $L$, the upper bound on the latency, $o_s$, the overhead of processor involving sending a message, $o_r$, the overhead of processor involving receiving a message, and $g$, the gap between consecutive message transmission. The measurement method relies on the Active Messages (AM) protocol \cite{eicken1992active} and consists of the following four communication experiments:

\begin{itemize}
  \item In the first experiment, the sender issues a small number of messages, $N_s$, consecutively without receiving any reply. The time of this experiment is measured on the sender side and estimated as $T_s = N_s \cdot o_s$. Thus, from this equation $o_s$ can be found as $o_s = T_s/N_s$.
  \item In the second experiment, the sender issues a large number of messages, $N_l$ ($N_l >> N_s$), consecutively. Time to send a message increases due to arriving replies during sending a message. When the capacity limit of the network is reached, the send request will eventually stall. Thus, the time to send $N_l$ messages in one direction can be estimated as $T_l = N_l \cdot g$, and $g$ is found from this linear equation as $g = T_l/N_l$. The time of this experiment is again measured on the sender side.
  \item The third experiment is designed to find $o_r$. The sender issues $N_l$ messages in one direction with $\bigtriangleup$ amount of time between messages. The delay $\bigtriangleup$ is introduced in order to make sure that the reply from the receiver has reached the sender side and therefore the time to process the reply by the sender can be accurately estimated as $o_r$. The time of this experiment is measured on the sender side and estimated as  $T^{'} =  N_l \cdot (o_s +  \bigtriangleup + o_r)$. Since $\bigtriangleup$ and $o_s$ are known, $o_r$ can be found from this linear equation as $o_r = T^{'}/N_l - o_s - \bigtriangleup$. 
  \item The fourth experiment performs a round-trip of a single message. The time of this experiment is measured on the sender side and estimated as $RTT = 2 \cdot(o_s + L + o_r)$. From this linear equation, $L$ can be found as $L = RTT/2 - o_s - o_r$.
\end{itemize}

Kielmann et al. \cite{kielmann2000fast} propose a method of measurement of parameters of the PLogP  (Parametrized LogP) model. PLogP defines its model parameters, except for latency L, as functions of message size. The method consists of the following four communication experiments: 

\begin{itemize}
  \item The first experiment is designed to measure $g(0)$. The sender sends $N$ consecutive zero-byte messages followed by a single empty reply from the receiver. Network saturation  is achieved by increasing the number of messages, $N$. It is assumed that when the network is saturated, the time $T$ to send a large number of zero-byte messages can be estimated as $T = N \cdot g(0)$, and $g$ can be found by solving this linear equation as $g(0) = T/N$. The time of this experiment is measured on the sender side.
  \item The second experiment is designed to measure $o_s(m)$. The sender starts the clock, sends a single message of size $m$, and then stops the clock. The time of this experiment is estimated as $T_s(m) = o_s(m)$.
  \item The third experiment is designed to measure $o_r(m)$. The sender sends a zero-byte message to the receiver,  waits for $\bigtriangleup$ time ($\bigtriangleup > T_s(m)$), starts the clock, receives a message of size $m$ and then stops the clock. The receiver receives the zero-byte message from the sender and sends back a message of size $m$. The time $T_r(m)$ measured on the sender side is estimated as $T_r(m) = o_r(m)$.
  \item The fourth experiment is designed to measure $L$ and $g(m)$. It consists of two round-trips,  with a zero-byte message and a message of size $m$ respectively. The time of the first round-trip is estimated as $RTT(0) = 2(L + g(0))$,  and the time of the second round-trip is estimated as $RTT(m) = 2\cdot L + g(0) + g(m)$. Both times are measured on the sender side. $L$ and $g(m)$ are then found from this system of two linear equations as $L = RTT(0)/2 - g(0)$ and $g(m) = RTT(m) - RTT(0) + g(0)$.  
\end{itemize}

Hoefler et al. \cite{hoefler2009loggp} develop a method to measure parameters of the LogGP model. LogGP extends the  LogP model by adding a $G$ parameter, the gap per byte for long messages. The building block of the method is a \textit{ping-ping} round-trip, where the sender sends $N$ consecutive messages of size $m$ with delay $d$ to the receiver, the receiver first receives all these messages and then sends them back to the sender, which also receives them all. The execution time of each communication experiment of the method, $PRTT(N,d,m)$, depends on parameters $N$, $d$ and $m$ of the experiment and measured on the sender side (PRTT stands for  Parametrized Round-Trip Time). Three particular ping-ping round-trip experiments are used to obtain equations involving the LogGP model parameters as unknowns:

\begin{itemize}
  \item The first experiment executes a round-trip of a single message $(N=1)$ of size $m$ without delay $(d = 0)$. The time of this experiment, $PRTT(1,0,m)$, is estimated as $PRTT(1,0,m) = 2 \cdot (o_s + L + o_r + (m - 1) \cdot G)$. 
  \item The second experiment executes a ping-ping round-trip that issues $N$ consecutive messages of size $m$ without delay $(d = 0)$. The time of this experiment, $PRTT(N,0,m)$, is estimated as $PRTT(N,0,m) = PRTT(1,0,m) + (N - 1) \cdot G_{all}$, where $G_{all}$ is a cumulative hardware gap, estimated as $G_{all} = G \cdot (m - 1) + g$.   
  \item The third experiment executes a ping-ping round-trip that issues $N$ consecutive messages of size $m$ with delay $d > 0$. The time of this experiment, $PRTT(N,d,m)$,  is estimated as $PRTT(N,d,m) = PRTT(1,0,m) + (N - 1) \cdot \max {\{o_s + d, G_{all}\}}$. 
\end{itemize}

Now model parameters $g$, $G$, $L$, $o_r$ and $o_s$ are found as follows:

\begin{itemize}
  \item From equations obtained from the first and second experiments, the  linear equation $G \cdot (m - 1) + g = \frac{PRTT(N,0,m) - PRTT(1,0,m)}{N - 1}$, involving two unknown parameters $g$ and $G$, can be derived. By repeating these experiment for a wide range of message size $m$, a system of $m$ linear equations with $g$ and $G$ as unknowns is produced. To find $g$ and $G$ from this system, the linear least-squares regression can be used.
  \item From the first experiment with $m=1$, the equation $PRTT(1,0,1) = 2 \cdot (o_s + L + o_r)$ can be derived, giving $L = PRTT(1,0,1)/2 - (o_s + o_r)$. However, the authors argue that due to the overlap of processor overheads and network latency, $L$ should be more accurately estimated as $L = PRTT(1,0,1)/2$.
  \item  In order to measure $o_r$, the measurement method proposed by Kielmann \cite{kielmann2000fast} is used.
  \item Finally, $o_s$ is found from the linear equation $o_s + d_G = \frac{PRTT(N,d_G,m) - PRTT(1,0,m)}{N - 1}$, which is derived from the first and third experiments, as $o_s = \frac{PRTT(N,d_G,m) - PRTT(1,0,m)}{N - 1} - d_G$. Here, parameter $d = d_G$ of the third experiment is determined empirically to guarantee that $d_G > G_{all}$.
\end{itemize}

\begin{table*}
  \centering
  \begin{tabular}{|l|l|l|}
    \hline
    \textbf{Collective routine} & \textbf{Collective algorithm}  \\ \hline \hline 
    
    Allgather & Ring \cite{thakur2005optimization}, Recursive doubling \cite{thakur2005optimization}, Bruck \cite{bruck1997efficient}, Neighbor exchange \cite{chen2005performance} \\[2ex] \hline 
    
    Broadcast & Flat tree \cite{pjesivac2007towards}, Chain tree \cite{patarasuk2006pipelined}, Binomial \cite{thakur2005optimization, pjesivac2007towards}, Binary \cite{pjesivac2007towards}, Split-binary \cite{pjesivac2007towards}, K-Chain tree \cite{pjesivac2007towards} \\[2ex] \hline 
    
    Barrier & Flat tree \cite{pjesivac2007towards}, Double Ring \cite{pjesivac2007towards}, Recursive doubling \cite{pjesivac2007towards}, Bruck \cite{bruck1997efficient} \\[2ex] \hline 
    
    Scater & Linear \cite{snir1998mpi}, Binomial \cite{pjesivac2007towards} \\[2ex] \hline 
    
    Gather & Linear \cite{pjesivac2007towards}, Linear with synchronisation \cite{pjesivac2007towards}, Binomial \cite{pjesivac2007towards} \\[2ex] \hline
    
    Alltoall & Linear \cite{pjesivac2007towards}, Pairwise exchange \cite{pjesivac2007towards}, Bruck \cite{bruck1997efficient} \\[2ex] \hline
    
    Reduce & Flat tree \cite{pjesivac2007towards}, Chain \cite{worringen2003pipelining, pjesivac2007towards}, Binomial \cite{pjesivac2007towards}, Binary \cite{pjesivac2007towards}, Rabenseifner \cite{rabenseifner1997new, rabenseifner2004more} \\[2ex] \hline  
    
    Reduce-scatter & Reduce-scatterv \cite{pjesivac2007towards}, Recursive halving \cite{pjesivac2007towards}, Ring \cite{pjesivac2007towards} \\[2ex] \hline
    
    Allreduce & Recursive doubling \cite{pjesivac2007towards}, Ring \cite{pjesivac2007towards}, Ring with segmentation \cite{pjesivac2007towards}, Rabenseifner \cite{rabenseifner1997new, rabenseifner2004more} \\[2ex] \hline
    
    Scan & Linear \cite{sanders2006parallel, pjesivac2007towards}, Linear with segmentation \cite{pjesivac2007towards}, Binomial \cite{pjesivac2007towards} \\[2ex] \hline

  \end{tabular} 
  \caption{List of collective algorithms used in Open MPI}  
  \label{tab:listofcollalgorithms}
\end{table*}

Rico-Gallego et al. \cite{rico2016extending} propose a detailed method for measurement of parameters of the $\tau$-Lop model on a multi-core cluster. $\tau$-Lop assumes that the cost of transmission of a message of size $m$ is estimated as $T^c_{p2p}(m) = o^c(m) + \sum_{j=0}^{s-1} L^c_{j}(m,\tau_{j})$, where $o^c(m)$ is the overhead of protocols and software stack, 
$ L^c_{j}(m,\tau_{j})$ is the time to transfer a message of size $m$ through channel $c$ at the $j$-th step of the transmission, with $\tau_{j}$ contending transfers ($ L^c_{j}(0,\tau_{j})=0$), and $s$ is the number of steps of the message transmission. For each communication channel, \textit{shared memory} or \textit{network}, experimental measurement of $o^c(m)$ is designed separately using the following round-trip experiments: 

\begin{itemize}
  \item The first experiment executes a round-trip of a message of size $m$ under the \textit{Eager} protocol for shared memory and network. The time of the experiment is estimated as $RTT^c(0) = 2 \cdot (o^c(m) + \sum_{j=0}^{s-1} L^c_{j}(0,1))$. For each channel, $o^c(m)$ is found as $o^c(m) = RTT^c(0)/2$.
  \item The second experiment executes a round-trip of a message of size $m$ under the \textit{Rendezvous} protocol for shared memory and network. The time of the experiment is estimated as $Ping^c(0) = o^c(m) + \sum_{j=0}^{s-1} L^c_{j}(0,1)$. 
  Therefore, $o^c(m) = Ping^c(0)$.
  \item The third set of experiments exchange messages of size $m$ between processes using \textit{MPI\_Sendrecv} routine in a ring shape. Process $P_i$ sends a message to $P_{i+1}$ and receives message from $P_{i-1}$. Then, \textit{MPI\_Wait} is called to complete both transmissions. $L^0$ and $L^1$ are estimated by the execution of these experiments in different channels respectively. 
\end{itemize}

From this overview, we can conclude that the state-of-the-art methods for measurement of parameters of communication performance models are all based on \textit{point-to-point communication experiments}, which are used to derive a system of equations involving model parameters as unknowns. In this work, we propose to use \textit{collective communication experiments} in the measurement method in order to improve the predictive accuracy of analytical models of collective algorithms. 

The only exception from this rule is a method for measurement of parameters of the LMO heterogeneous communication model  \cite{lastovetsky2007building,lastovetsky2009accurate}. LMO is a communication model of heterogeneous cluster, and the total number of its  parameters is significantly larger than the maximum number of independent point-to-point communication experiments that can be designed  to derive a system of independent linear equations with the model parameters as unknowns. To address this problem and obtain the sufficient number of independent linear equations involving model parameters, the method additionally introduces simple collective communication experiments, each using three processors and consisting of a one-to-two communication operation (scatter) followed by a two-to-one communication operation (gather). The experiments are implemented using the MPIBlib library \cite{lastovetsky2008mpiblib}. This method however is not designed to improve the accuracy of predictive analytical models of communication algorithms.

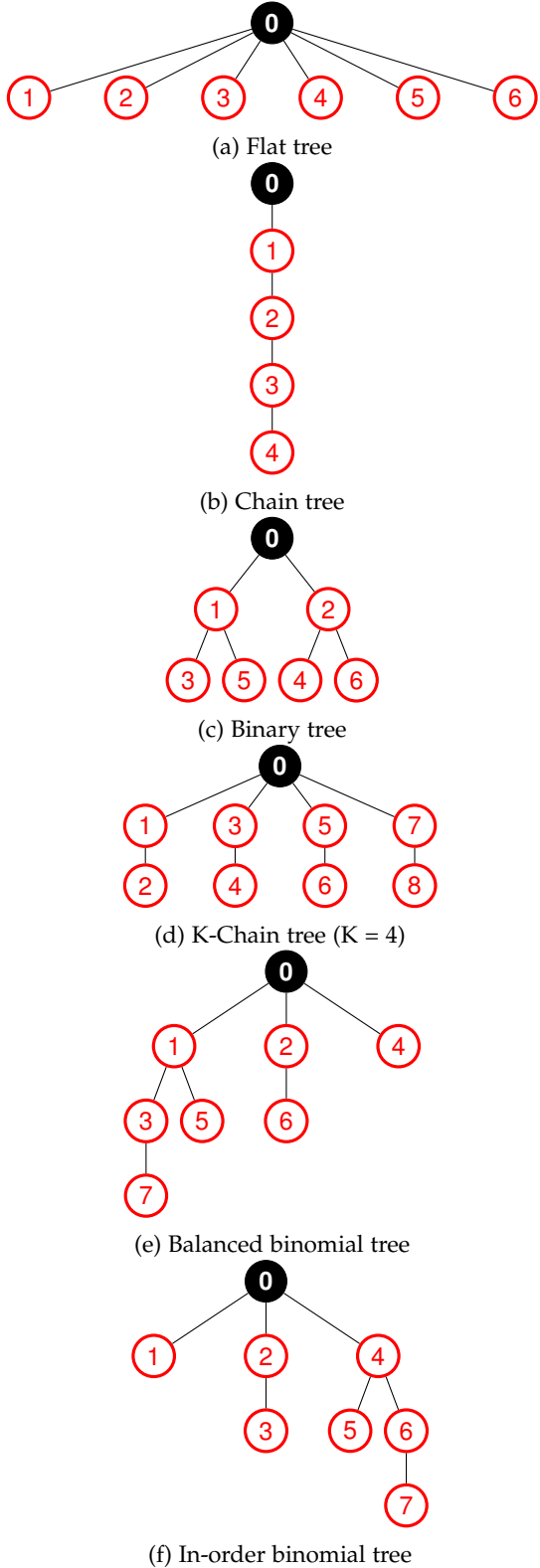
\begin{figure}[t!]
  \centering
  \begin{subfigure}[t]{0.4\textwidth}
  \centering
  \vspace{1.1cm}
  \begin{tikzpicture}[>=stealth',level/.style={sibling distance = 1.3cm/#1,
    level distance = 1cm}] 
  \node [root_style] {0}
      child{ node [child_style] {1} }
      child{ node [child_style] {2} }
      child{ node [child_style] {3} }            
      child{ node [child_style] {4} }            
      child{ node [child_style] {5} }            
      child{ node [child_style] {6} }            
  ;
  \end{tikzpicture}
  \caption{Flat tree}
  \label{fig:flattreetop}
  \end{subfigure}

  \begin{subfigure}{.4\textwidth}
    \centering
    \begin{tikzpicture}[>=stealth',level/.style={sibling distance = 1cm/#1,
      level distance = 0.9cm}] 
    \node [root_style] {0}
        child{ node [child_style] {1} 
                child{ node [child_style] {2} 
                   child{ node [child_style] {3} 
                        child{ node [child_style] {4} }  
                     }
                }      
        }
    ;
    \end{tikzpicture}
    \caption{Chain tree}
    \label{fig:pipelinedtreetop}
  \end{subfigure}

  \begin{subfigure}[t]{.4\textwidth}
  \centering
  \begin{tikzpicture}[>=stealth',level/.style={sibling distance = 1.5cm/#1,
    level distance = 0.95cm}] 
  \node [root_style] {0}
      child{ node [child_style] {1} 
              child{ node [child_style] {3} }
              child{ node [child_style] {5} }                            
      }
      child{ node [child_style] {2}
              child{ node [child_style] {4} }
              child{ node [child_style] {6} }
      }
  ;
  \end{tikzpicture}
  \caption{Binary tree }
  \label{fig:binarytreetop}
  \end{subfigure}
  \begin{subfigure}[t]{.5\textwidth}
  \centering
  \vspace{0cm}
  \begin{tikzpicture}[>=stealth',level/.style={sibling distance = 1.2cm/#1,
    level distance = 0.8cm}] 
  \node [root_style] {0}
      child{ node [child_style] {1} 
        child { node [child_style] {2}}
      }
      child{ node [child_style] {3} 
        child { node [child_style] {4}}
      }
      child{ node [child_style] {5} 
        child { node [child_style] {6}}
      }            
      child{ node [child_style] {7} 
        child { node [child_style] {8}}
      }            
  ;
  \end{tikzpicture}
  \caption{K-Chain tree (K = 4)}
  \label{fig:kchaintop}
  \end{subfigure}
    
  \begin{subfigure}{.4\textwidth}
  \centering
  \begin{tikzpicture}[>=stealth',level/.style={sibling distance = 1.5cm/#1,
    level distance = 1cm}] 
  \node [root_style] {0}
      child{ node [child_style] {1}
        child { node [child_style] {3}
            child { node [child_style] {7}}
          }
          child { node [child_style] {5}}
      } 
      child{ node [child_style] {2}
        child { node [child_style] {6}}
      }
      child{ node [child_style] {4}}
  ;
  \end{tikzpicture}
  \caption{Balanced binomial tree}
  \label{fig:binomialtreebaltop}
  \end{subfigure}
  \begin{subfigure}{.5\textwidth}
    \centering
    \begin{tikzpicture}[>=stealth',level/.style={sibling distance = 1.5cm/#1,
      level distance = 1cm}] 
    \node [root_style] {0}
        child{ node [child_style] {1}} 
        child{ node [child_style] {2}
          child { node [child_style] {3}}
        }
        child{ node [child_style] {4}
          child { node [child_style] {5}}
          child { node [child_style] {6}
            child { node [child_style] {7}}
          }
        }
    ;
    \end{tikzpicture}
    \caption{In-order binomial tree}
    \label{fig:binomialtreeintop}
    \end{subfigure}  
  \caption{\textit{Virtual topologies for collective algorithms}}
  \label{fig:virtualtopologies}
\end{figure}

\section{Collective algorithms}
\label{sec:collalgorithms}

In this work, we propose to derive analytical models of MPI collective algorithms from their implementations rather than from  high-level mathematical definitions, and  use the derived  models at runtime for selection of the optimal  algorithms. We present this approach by applying it  to Open MPI and its  \textit{broadcast} and \textit{gather} collective algorithms. 
While the algorithms are Open MPI specific, the proposed modelling approach itself is general and can be applied to other MPI implementations and collective algorithms.

\subsection{Open MPI \textit{broadcast} and \textit{gather} collective algorithms}

The complete list of Open MPI 3.1 collective algorithms  can be found  in Table \ref{tab:listofcollalgorithms}. In this work, we cover in detail the \textit{broadcast} and \textit{gather} algorithms.
 
During the broadcast (\textit{MPI\_Bcast}), one process, called \textit{root}, sends the same data to all processes in the communicator. At the end of the operation, the root buffer is copied to all other processes. Let  $P$ be the number of processes involved in the collective operation. All broadcast algorithms used in Open MPI implementation are listed below:

\begin{itemize}
\item \textbf{Linear tree algorithm}. The algorithm employs a single level tree topology shown in Figure \ref{fig:flattreetop} where the root node has $P - 1$ children. The message is transmitted to child nodes without segmentation.

\item \textbf{Chain tree algorithm}. Each internal node in the topology has one child (see Fig \ref{fig:pipelinedtreetop}). The message is split into segments and transmission of segments continues until last node gets the broadcast message.

\item \textbf{Binary tree algorithm}. Unlike the chain tree, each internal process has two children where data is distributed from root to all leaves (Figure \ref{fig:binarytreetop}). Segmentation technique is employed in this algorithm.

\item \textbf{Split binary tree algorithm}. The split binary tree algorithm employs the same virtual topology as the binary tree (Figure \ref{fig:binarytreetop}). As the name implies, the difference from the binary tree algorithm is the splitting of the message into two halves before transmission. After splitting the message, the right and left halves of the message are pushed down into the right and left subtrees respectively. In the last phase, the left and right processes exchange their halves of the message to complete the broadcast operation.

\item \textbf{K-Chain tree algorithm}. In the \textit{K-Chain tree} topology (Figure \ref{fig:kchaintop}), the root process has $K$ children and each internal process in the tree has only one child. The root broadcasts the message to the child processes, then the child processes broadcast the message to their children in parallel. The height of \textit{K-chain tree} is estimated as $H_{k-chain} = \lfloor \frac{P - 1}{K} \rfloor$. Last process must wait for $H_{k-chain}$ steps until it gets the broadcast message. The algorithm is implemented using the segmentation technique.

\item \textbf{Binomial tree algorithm}. The binomial tree topology is determined according to the binomial tree definition \cite{huse1999collective}. The algorithm employs \textit{balanced} binomial tree (Figure \ref{fig:binomialtreebaltop}). Unlike the binary tree, the maximum nodal degree of the binomial tree decreases from the root down to the leaves as follows: $\lceil \log_2 P \rceil, \lceil \log_2 P \rceil - 1, \lceil \log_2 P\rceil - 2, ...$. The height of the binomial tree is the order of the tree, $H = \lfloor \log_2 P \rfloor$.

\label{list:broadcast_algorithms}
\end{itemize}

\textit{MPI\_Gather} is a popular \textit{many-to-one} MPI operation. MPI\_Gather takes data elements from all processes of the communicator and gathers them in one single process which is called \textit{root}. MPI\_Gather is used in many parallel applications such as parallel sorting and searching. The complete list of gather algorithms employed in Open MPI is as follows:

\begin{itemize}
\item \textbf{Linear algorithm without synchronisation}. This algorithm employs flat tree virtual topology (Figure \ref{fig:flattreetop}). In this algorithm, the non-root processes send their messages to the root, which posts receives from everyone. If the operation is not denoted as “in-place”,  the root must perform a local copy of its own data.
\item \textbf{Linear algorithm with synchronisation}. The algorithm uses flat tree virtual topology (Figure \ref{fig:flattreetop}) as well. This algorithm was introduced to prevent overloading of the root process using message segmentation technique. The message is split into two segments on each non-root process. To receive the message from non-root processes, the algorithm performs the following steps: (1) The root receives the first incoming segment of the message; (2) Then, the root sends a zero-byte message to non-root processes, signalling them to send the second segment of the message; (3) The root receives the second segment of the message. 
\item \textbf{Binomial algorithm}. The binomial algorithm employs in-order binomial tree topology (Figure \ref{fig:binomialtreeintop}). 
The leaf nodes send their data to their parent processes immediately. Internal nodes in the tree wait to receive the data from all children before forwarding the message up the tree. Once the root node receives all messages, a local data shift operation may be necessary to put the data in the correct place. 
\label{list:gather_algorithms}
\end{itemize}

\renewcommand{\algorithmiccomment}[1]{// #1}
\begin{algorithm}[t]
\begin{algorithmic}
\caption{Tree-based segmented broadcast algorithm}
\label{alg:openmpibcastdesign}
      \IF {$(rank == root)$}
      \STATE \COMMENT{Send segments to all children}
		\FOR{$i \in 0 .. n_{s}-1$}		
			\FOR {$child$ $\in$ $list$ $of$ $children$}
				\STATE MPI\_Isend(segment[$i$], child, ... )
			\ENDFOR
			\STATE MPI\_Waitall(. . .)			
		\ENDFOR        
      \ELSIF {$(intermediate$ $nodes)$}
		\STATE \COMMENT{Post receive and wait}    
        \STATE MPI\_Irecv($segment$)  
        \STATE MPI\_Wait(. . .)
		\STATE \COMMENT{Send data to children}         	
        	\FOR {$child in list of children$}
			\STATE MPI\_Isend($segment, child$, ... )
		\ENDFOR
        	\STATE MPI\_Waitall($children$)	
	  \ELSIF {$(leaf$ $nodes)$}
		\STATE \COMMENT{Receive all segments from parent in a loop}	    
	    \FOR{$i \in 0 .. n_{s}-1$}		
	    		\STATE MPI\_Irecv($segment$, ... )   
	    		\STATE MPI\_Wait(. . .)  
	    	\ENDFOR
      \ENDIF    

\end{algorithmic}
\end{algorithm}

\section{Implementation-derived analytical models of broadcast and gather algorithms}
\label{sec:implementationdrivenmodel}

As stated  in Section \ref{sec:intro}, we propose a new approach to analytical performance modelling of collective algorithms. While the traditional approach only takes into account high-level mathematical definitions of the algorithms, we derive our models from their implementation. This way, our models take into account important details of their execution having a significant impact on their performance. In this section, we present our analytical modelling approach by applying it to broadcast and gather collective algorithms implemented in Open MPI. This approach could be similarly applied to other  collective algorithms 
and  MPI implementations such as MPICH. 
Analytical models of the broadcast and gather collective algorithms implemented in Open MPI are derived in Sections \ref{subsec:bcastmodelling} and \ref{subsec:gathermodelling}. 

To model point-to-point communications, we use the Hockney model, which estimates the time $T_{p2p}(m)$ of sending a message of size $m$ between two processes as $ T_{p2p}(m) = \alpha + \beta \cdot m $, where $\alpha$ and  $\beta$ are the latency and the reciprocal bandwidth respectively. For segmented collective algorithms, we assume that $m = n_s \cdot m_s$, where $n_s$ and $m_s$ are the number of segments and the segment size respectively. We assume that each algorithm involves $P$ processes ranked from $0$ to $P - 1$.

\subsection{Broadcast algorithms}
\label{subsec:bcastmodelling}

In this section, we build analytical performance models of broadcast algorithms implemented in Open MPI. All broadcast algorithms implemented in Open MPI, except for the linear tree broadcast algorithm, are implemented using message segmentation. As the main purpose of message segmentation is to avoid the \textit{rendezvous} protocol, we only build analytical models of broadcast algorithms with message segmentation assuming the buffered mode of \textit{send} operations. Models of segmented broadcast algorithms employing the \textit{rendezvous} (synchronous) mode would have no practical application in Open MPI as they assume a configuration with a segment size being not small enough to avoid the \textit{rendezvous} protocol, which does not make much sense.

\subsubsection{Linear (Flat) tree algorithm}
\label{subsec:analyticalmodellineartreebcast}

In Open MPI, the linear broadcast algorithm is implemented using blocking \textit{send} and \textit{receive} operations. The algorithm transmits the whole message from root to the leaves without message segmentation. Regardless of communication mode (buffered or not), because of blocking communication, each next \textit{send} only starts after the previous one has been completed. Therefore, the execution time of the linear tree broadcast algorithm will be equal to the sum of execution times of $P - 1$ send operations:

\begin{equation}
  T^{blocking}_{linear\_bcast}(P,m) = (P - 1) \cdot (\alpha + m \cdot \beta).
  \label{equ:lineartreebcastblocking}
\end{equation}

In Open MPI, this linear tree algorithm is one of the six algorithms available for implementation of the \textit{MPI\_Bcast} routine. There is another linear tree broadcast algorithm, which cannot be chosen to implement \textit{MPI\_Bcast}, but only used as a building block in other tree-based broadcast algorithms implementing \textit{MPI\_Bcast}, namely, in the  \textit{binomial tree}, \textit{binary tree}, \textit{k-chain tree}, and \textit{chain tree} broadcast algorithms (see Algorithm \ref{alg:openmpibcastdesign} for more details). That linear tree algorithm is implemented using \textit{non-blocking} send and receive operations. 
 
In this latter case, $P-1$ non-blocking \textit{send}s will run on the \textit{root} concurrently. Therefore, the execution time of the linear broadcast algorithm using non-blocking point-to-point communications and buffered mode, $T^{nonblock}_{linear\_bcast}(P, m)$, can be bounded as follows:

\begin{equation}
  T_{p2p}(m) \leq T^{nonblock}_{linear\_bcast}(P, m) \leq (P - 1) \cdot T_{p2p}(m).
  \label{equ:definitionbufnonlinearbcast}
\end{equation}

\noindent We will approximate $T^{nonblock}_{linear\_bcast}(P, m)$ as 

\begin{equation}
  T^{nonblock}_{linear\_bcast}(P, m) = \gamma (P,m) \cdot (\alpha + m \cdot \beta),
  \label{equ:costoflinearbcastnonbuffered}
\end{equation} 

\noindent where 

\begin{equation}
  \gamma (P,m) = \frac{T^{nonblock}_{linear\_bcast}(P, m)}{T_{p2p}(m)}.  
  \label{equ:gammafunction}
\end{equation} 

We will use this approximation when deriving analytical performance models of the remaining five broadcast algorithms implemented in Open MPI. As we can see from Algorithm \ref{alg:openmpibcastdesign}, the non-blocking version of linear tree broadcast is used in these five algorithms for transmission of a single message segment. In this paper, we assume the same fixed  segment size in all segmented  algorithms. Therefore, in the rest of the paper we define $\gamma$ as a function of $P$ only, $\gamma(P)$. From Formula \ref{equ:definitionbufnonlinearbcast}, we can derive that $T^{nonblock}_{linear\_bcast}(2, m) = T_{p2p}(m)$ and, hence,  $\gamma(2) = 1$.

\begin{figure}[t]
	\centering
	\includegraphics[width=6cm]{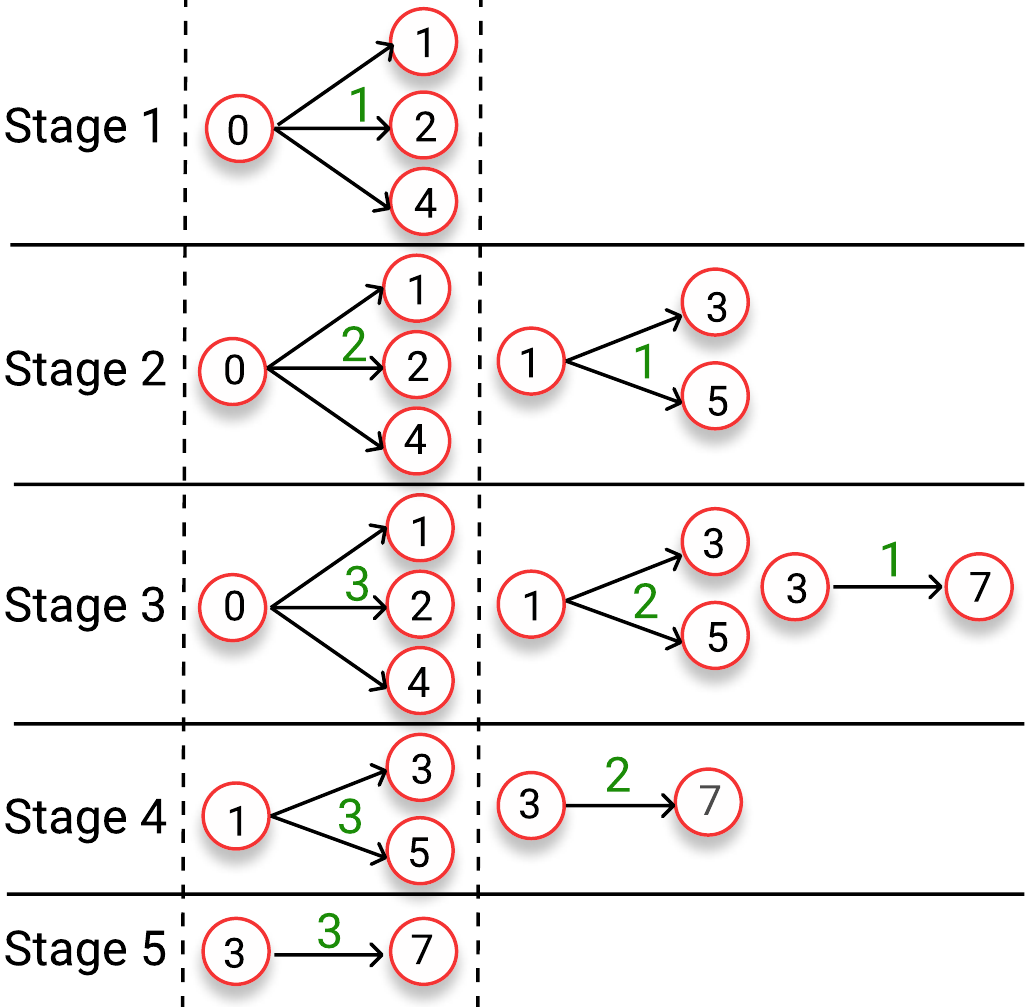}
  \caption{ Execution stages of the binomial broadcast algorithm, employing the non-blocking linear broadcast
($P = 8$, $ns = 3$). Nodes are labelled by the process ranks. Each arrow represents transmission of a segment. The number over the arrow gives the index of the broadcast segment.}
  \label{fig:executionstepsbinomialbcast}
\end{figure}

\subsubsection{Binomial tree algorithm}

In Open MPI, the binomial tree broadcast algorithm is segmentation-based and implemented as a combination of linear tree broadcast algorithms using non-blocking \textit{send} and \textit{receive} operations. 

Figure \ref{fig:executionstepsbinomialbcast} shows the stages of execution of the binomial tree broadcast algorithm. Each stage consists of parallel execution of a number of linear broadcast algorithms using non-blocking communication.The linear broadcast algorithms running in parallel have a different number of children. Therefore, the execution time of each stage will be equal to the execution time of the linear broadcast algorithm with the maximum number of children. The execution time of the whole binomial broadcast algorithm will be equal to the sum of the execution times of these stages. 

In Open MPI, the binomial tree broadcast algorithm employs the balanced binomial tree virtual topology. Therefore, the number of stages in the binomial broadcast algorithm can be calculated as

\begin{equation}
  N_{steps} = \lfloor log_2 P \rfloor + n_s - 1.
  \label{equ:generalstepstreebasedbcast}
\end{equation}

Thus, the time to complete the binomial tree broadcast algorithm can be estimated as follows:
\begin{equation}
   \begin{aligned}[b]
  T_{binomial\_bcast}(P,m,n_s) = \\
\sum_{i=1}^{\lfloor log_2 P \rfloor + n_s - 1} \max_{1 \leq j \leq \min (\lfloor log_2 P \rfloor, n_s)} T^{nonblock}_{linear\_bcast}(P_{ij},\frac{m}{n_s}), 
\label{equ:generalcostbinomialtreebcast}
   \end{aligned}
\end{equation}
\noindent where $P_{ij}$ denotes the number of nodes in the $j$-th linear tree of the $i$-th stage.

Using the property of the binomial tree and Formula \ref{equ:costoflinearbcastnonbuffered}, we have 
\begin{equation}
   \begin{aligned}[b]
    T_{binomial\_bcast}(P,m,n_s) = (n_s \cdot \gamma(\lceil \log_2 P \rceil+1) + \\ \sum_{i=1}^{\lfloor \log_2 P \rfloor - 1}\gamma (\lceil \log_2 P \rceil - i + 1) - 1) \cdot (\alpha + \frac{m}{n_s} \cdot \beta).
   \end{aligned}
\end{equation}


\subsubsection{Chain tree algorithm} 

In Open MPI, the chain tree algorithm is segmentation-based and implemented using non-blocking point-to-point communication. While the height of the chain tree equal to $P - 1$, the algorithm will be completed in $P + n_s - 2$ steps, each  consisting of a varying number of concurrent non-blocking point-to-point communications (technically, Open MPI employs concurrent non-blocking linear tree broadcast algorithms, but in this case each linear broadcast will be equivalent to a point-to-point communication). 
Therefore, the execution time of the chain tree algorithm can be estimated as

\begin{equation}
  T_{chain\_bcast}(P,m,n_s) = (P + n_s - 2) \cdot (\alpha + \frac{m}{n_s} \cdot \beta). 
\end{equation}

\subsubsection{Split-binary tree algorithm}

In Open MPI, the split-binary tree algorithm is segmentation-based and implemented using blocking point-to-point communication. The algorithm consists of two phases -- \textit{forwarding} and \textit{exchange}. In the first phase, the message of size $m$ is split into two equal parts in the root, which are then sent to the left and right subtrees respectively using message segmentation. After completion of the first phase, each node in the left subtree contains the first half of the message and each node in the right subtree -- the second half of the message. Because of segmentation, each node will receive $\frac{n_s}{2}$ segments during the first phase. 

As the balanced binary tree virtual topology is employed  in the split-binary tree algorithm, each node in the left subtree will have a matching pair in the right subtree and vice versa. In the second phase, each pair of matching nodes in the left and right subtrees exchange their halves of the message. The execution time of the split-binary tree broadcast will be equal to the sum of the execution times of the first and the second phases. As the heigh of the balanced binary tree is equal to $\lfloor \log_2 P \rfloor$, we have

\begin{equation}
  \begin{aligned}[b]
    T_{split\_binary\_bcast}(P,m,n_s) = 2 \cdot (\lfloor \log_2 P \rfloor + \frac{n_s}{2} - 1) \cdot \\ (\alpha + \frac{m}{n_s} \cdot \beta) + (\alpha + \frac{m}{2} \cdot \beta)     
  \end{aligned}
\end{equation}

\subsubsection{Binary tree algorithm} 

In Open MPI, the  binary tree broadcast algorithm is segmentation-based and uses the balanced binary tree topology (see Figure \ref{sub@fig:binarytreetop}). The root broadcasts each segment to its children using the non-blocking  linear tree broadcast algorithm. Upon receipt of next segment, each internal node acts similarly.  
As the binary tree used in this algorithm is balanced, all the non-blocking linear broadcasts will have the same execution time, namely,\\ 

$ T^{nonblock}_{linear\_bcast}(3, m_s) = \gamma(3) \cdot (\alpha + \frac{m}{m_s} \cdot \beta)$. \\

\noindent As the height of the balanced binary tree is equal to $\lfloor \log_2 P \rfloor$, the algorithm will be completed in $(\lfloor \log_2 P \rfloor + n_s - 1)$ steps, each  consisting of a varying number of concurrent non-blocking linear broadcasts, involving $3$ processes. Therefore, 

\begin{equation}
  \begin{aligned}[b]
    T_{binary\_bcast}(P,m,n_s) = \gamma(3) \cdot (\lfloor \log_2 P \rfloor + n_s - 1) \cdot \\ (\alpha + \frac{m}{n_s} \cdot \beta).   
  \end{aligned}  
\end{equation}

\subsubsection{K-chain tree algorithm}
  
In Open MPI, the K-chain tree algorithm is implemented using non-blocking communication and message segmentation. In the K-chain tree, the root node has $K (K > 1)$ children, while the internal nodes have a single child each (Figure \ref{sub@fig:kchaintop}). As the height of the tree is $\lfloor \frac{P - 1}{K} \rfloor$, 
the algorithm takes $\lfloor \frac{P - 1}{K} \rfloor + n_s - 1$ steps to complete. At each step, a varying number ofl non-blocking linear tree broadcast algorithms will be executed concurrently (one at the first step, $K$ at the last step, and up to $K \times (\lfloor \frac{P - 1}{K} \rfloor -1) + 1$  algorithms for intermediate steps). Note, that while Open MPI employs concurrent non-blocking linear tree broadcast algorithms, in this case the most of the linear broadcasts will be equivalent to non-blocking point-to-point communications.

The execution time of the K-chain tree algorithm will be equal to the sum of the execution times of its steps. 
The execution time of each step will be equal to the maximum execution time of the concurrently executed linear broadcasts. 
For the first $n_s$ steps, this maximum time will be the time of the linear broadcast involving the root of the whole K-chain tree, which is estimated as 
$\gamma(K+1) \cdot (\alpha + \frac{m}{n_s} \cdot \beta)$ according to Formula \ref{equ:costoflinearbcastnonbuffered}. For each of the remaining $\lfloor \frac{P - 1}{K} \rfloor - 1$ steps, all concurrently executed linear broadcasts will be equivalent to non-blocking point-to-point communications, the time of which is $\alpha + \frac{m}{n_s} \cdot \beta$. Thus, the total execution time of the K-chain tree algorithm will be estimated as

\begin{equation}
  \begin{aligned}[b]
    T_{k\_chain\_bcast}(P,m,n_s) = (\lfloor \frac{P - 1}{K} \rfloor + \\ \gamma(K+1) \cdot n_s - 1) \cdot (\alpha + \frac{m}{n_s} \cdot \beta).
  \end{aligned}  
\end{equation}

\begin{figure*}
  \begin{subfigure}[t]{0.33\textwidth}
    \centering
    \resizebox{\linewidth}{!}{
              \begin{tikzpicture}
          \begin{axis}[
              xmode=log,
              log basis x={2},
              title={},
              xlabel={Message size},
              ylabel={Time(sec)},
               legend style={
                              at={(0,0.8)},
                              anchor=south west,
                              /tikz/every even column/.append style={column sep=1.0cm}
                                  }
          ]
           
          \addplot[
              color=blue,
              mark=square,
              ]
              coordinates {
(8192,0.00047580888556230955)
(16384,0.000951617764249048)
(32768,0.0019032355216225248)
(65536,0.0038064710363694782)
(131072,0.007612942065863386)
(262144,0.015225884124851199)
(524288,0.030451768242826832)
(1048576,0.06090353647877809)
(2097152,0.12180707295068059)
(4194304,0.24361414589448566)
              };
           \addlegendentry{Binary tree}   
        
           \addplot[
              color=red,
              mark=triangle,
              ]
              coordinates {
(8192,0.0002775551832446806)
(16384,0.0005551103624786113)
(32768,0.0011102207209464727)
(65536,0.002220441437882196)
(131072,0.004440882871753642)
(262144,0.008881765739496533)
(524288,0.017763531474982316)
(1048576,0.03552706294595388)
(2097152,0.07105412588789702)
(4194304,0.1421082517717833)
              };
            \addlegendentry{Binomial tree}       
          \end{axis}
          \end{tikzpicture}
    }
    \caption{} 
    \label{subfig:binomialbinaryexist}
  \end{subfigure} %
  \begin{subfigure}[t]{0.33\textwidth}
    \centering
    \resizebox{\linewidth}{!}{
              \begin{tikzpicture}
          \begin{axis}[
              xmode=log,
              log basis x={2},
              title={},
              xlabel={Message size},
              ylabel={Time(sec)},
              legend style={
                              at={(0, 0.8)},
                              anchor = south west,
                              /tikz/every even column/.append style={column sep=1.0cm}
                                  }
          ]
           
          \addplot[
              color=blue,
              mark=square,
              ]
              coordinates {
(8192, 0.0015113012)
(16384, 0.0016055487)
(32768, 0.0017970541)
(65536, 0.0025138593)
(131072, 0.0033746346)
(262144, 0.0049619167)
(524288, 0.0083341763)
(1048576, 0.0147779914)
(2097152, 0.0276803771)
(4194304, 0.0525307779)
              };
           \addlegendentry{Binary tree}   
        
           \addplot[
              color=red,
              mark=triangle,
              ]
              coordinates {
(8192, 0.0015266695)
(16384, 0.001598808)
(32768, 0.001748272)
(65536, 0.0029432779)
(131072, 0.0046290569)
(262144, 0.0074023664)
(524288, 0.0134850847)
(1048576, 0.0266538259)
(2097152, 0.0526081832)
(4194304, 0.0990790485)     
              };
            \addlegendentry{Binomial tree}      
          \end{axis}
          \end{tikzpicture}
    }
    \caption{}
    \label{subfig:binomialbinarexp}%
  \end{subfigure}%
  \begin{subfigure}[t]{0.33\textwidth}
    \centering
    \resizebox{\linewidth}{!}{
              \begin{tikzpicture}
          \begin{axis}[
              xmode=log,
              log basis x={2},
              title={},
              xlabel={Message size},
              ylabel={Time(sec)},
               legend style={
                              at={(0,0.8)},
                              anchor=south west,
                              /tikz/every even column/.append style={column sep=1.0cm}
                                  }
          ]
           
          \addplot[
              color=blue,
              mark=square,
              ]
              coordinates {
(8192,0.00047580888556230955)
(16384,0.000951617764249048)
(32768,0.0019032355216225248)
(65536,0.0038064710363694782)
(131072,0.007612942065863386)
(262144,0.015225884124851199)
(524288,0.030451768242826832)
(1048576,0.06090353647877809)
(2097152,0.12180707295068059)
(4194304,0.24361414589448566)
              };
           \addlegendentry{Binary tree}   

		 \addplot[
              color=red,
              mark=square,
              ]
              coordinates {
(8192,0.0008326655497340417)
(16384,0.0016653310874358341)
(32768,0.003330662162839418)
(65536,0.006661324313646587)
(131072,0.013322648615260925)
(262144,0.0266452972184896)
(524288,0.053290594424946956)
(1048576,0.10658118883786165)
(2097152,0.21316237766369103)
(4194304,0.42632475531534986)
              };
           \addlegendentry{Binomial tree}          
               
          \end{axis}
          \end{tikzpicture}
    }
    \caption{}
    \label{subfig:binomialbinarour}
  \end{subfigure}

  \caption{Performance estimation of the binary and binomial tree broadcast algorithms  by the traditional  and proposed analytical  models in comparison with experimental curves. The experiments involve ninety processes (P=90). (\subref{subfig:binomialbinaryexist}) Estimation by the existing analytical models. (\subref{subfig:binomialbinarexp}) Experimental performance curves. (\subref{subfig:binomialbinarour}) Estimation by the proposed analytical models derived from the implementation codes.}%
  \label{fig:binarybinomialperformance}
\end{figure*}

\subsection{Gather algorithms}
\label{subsec:gathermodelling}

In this section, we derive analytical formulas of the gather algorithms implemented in Open MPI.

\subsubsection{Linear without synchronisation}

In the Open MPI implementation of the \textit{linear without synchronisation} gather  algorithm, 
the root receives messages from its $P-1$ children using blocking receive operations. 
Therefore, the execution time of this gather algorithm can be be estimated as the sum of the execution times of $P - 1$ blocking receive operations, that is,

\begin{equation}
  T_{linear\_gather}(P,m) = (P - 1) \cdot (\alpha + m \cdot \beta).
\end{equation}

\subsubsection{Linear with synchronisation}

The Open MPI implementation of the \textit{linear with synchronisation} gather algorithm employs both blocking and non-blocking communications. 
The messages gathered from the children are all identically split into two equal parts.
In order to receive all these parts from its $P-1$ children, the root executes a loop, at $i$-th iteration of which it receives both halves of the message from the $i$-th child by performing
the following steps: 1) it first posts a non-blocking receive for the first part; 2) then it sends a zero-byte message using a blocking send, signalling the child  to start sending the message parts; 3) then the root posts a non-blocking receive for the second half of the message; 4) finally, it blocks itself waiting for the completion of the previously posted non-blocking receives.

At the same time, upon receipt of a zero-byte signal message from the root, each child will perform two successive standard blocking sends for the first and the second parts of its message. When the size of these parts, $\frac{m}{2}$, is grater than the eager limit, $m_{eager}$, than the standard blocking sends will follow the \textit{rendesvouz} protocol, that is, will be equivalent to \textit{synchronous} sends. Otherwise, they will follow the \textit{eager} protocol, that is, will be equivalent to \textit{buffered} sends. In the first case, the execution of all point-to-point communications will be serialized, 
and, therefore, the execution time of the linear gather with synchronisation algorithm can be estimated as the sum of the execution times of the employed point-to-point communications: 

\begin{equation}
   \begin{aligned}[b]
  T_{linear\_gather\_with\_synch}(P,m) = (P - 1) \cdot (2 \cdot (\alpha + \frac{m}{2} \cdot \beta))\\ = (P - 1) \cdot (2 \cdot \alpha + m \cdot \beta).
   \end{aligned}
\end{equation}

\noindent Otherwise, when $\frac{m}{2} \leq m_{eager}$, each child will send its half-messages concurrently.  Therefore, the execution time of the linear gather with synchronisation algorithm in this case can be estimated as 

\begin{equation}
   \begin{aligned}[b]
  T_{linear\_gather\_with\_synch}(P,m) = (P - 1) \cdot (\alpha + \frac{m}{2} \cdot \beta).
   \end{aligned}
\end{equation}
\subsubsection{Binomial algorithm}

In Open MPI, the binomial gather algorithm employs the \textit{in-order} binomial tree virtual topology (Figure \ref{fig:binomialtreeintop}). The leaf nodes and internal nodes use the standard blocking send to send the messages to their parents, which receive the message using the blocking receive. The algorithm will be completed in $\lfloor \log_2 P \rfloor$ steps, each performing a set of concurrent blocking receives.

 At $i$-th step, the root will receive a message of size $2^{i-1} \cdot m$ from its $i$-th child, combining the messages gathered by the latter acting as the root of the $i$-th subtree during the previous $i-1$ steps ($i = 1,...,\lfloor \log_2 P \rfloor$).  Given this message size, $2^{i-1} \cdot m$, will be the largest communicated at the $i$-th step of the algorithm, its execution time can be estimated as


\begin{equation}
  \begin{aligned}[b]
    T_{binomial\_gahter}(P,m) = \sum_{i=1}^{\lceil \log_2 P \rceil }(\alpha + 2^{i-1} \cdot m \cdot \beta) \\= \lceil \log_2 P \rceil \cdot \alpha + (P - 1) \cdot m \cdot \beta.
  \end{aligned}  
\end{equation}    

\subsection{Comparison of the proposed analytical models against the state of the art}

In this section, we use the binomial and binary tree algorithms as an example to  illustrate that unlike the traditional approaches, the approach based on the derivation of analytical models of collective algorithms from their implementation codes, yields models, which can be used for accurate  pairwise comparison of the performance of  collective algorithms implementing the same collective operation. 

Existing analytical modelling approaches \cite{pjesivac2007towards,thakur2005optimization,hasanov2017hierarchical} 
estimate the execution time of the  binary  and binomial tree broadcast algorithms as follows:

\begin{center}
$T_{binomial\_bcast}(P, m) = \lceil \log_2 P \rceil \cdot (\alpha + m \cdot \beta),$
\end{center} 
 
\begin{center}
$T_{binary\_bcast}(P, m) = 2 \cdot (\lceil \log_2 (P + 1) \rceil - 1) \cdot (\alpha + m \cdot \beta).$
\end{center}

Figure \ref{fig:binarybinomialperformance} shows the performance of the binary tree and binomial tree algorithms using: a) the estimation by the existing analytical models; b) the experimental results on the Grisou cluster of the Grid'5000 platform; c) the estimation by the analytical models presented in Sectioin \ref{subsec:bcastmodelling}. It is evident that while the existing models wrongly predict that the binomial tree algorithm will outperform the binary tree algorithm on the target platform, our models correctly  predict the relative performance of these algorithms.







\begin{figure*}
  \centering
  \begin{flalign*}
      & \begin{cases}
      (\frac{P - 1}{K} + \gamma(K+1) \cdot n_{s_1} - 1) \cdot (\alpha + \frac{m_{1}}{n_{s_1}} \cdot \beta) + (P - 1) \cdot (\alpha + \frac{m_{1}}{n_{s_1}} \cdot \beta) = T_{1} \\
      (\frac{P - 1}{K} + \gamma(K+1) \cdot n_{s_2} - 1) \cdot (\alpha + \frac{m_{2}}{n_{s_2}} \cdot \beta) + (P - 1) \cdot (\alpha + \frac{m_{2}}{n_{s_2}} \cdot \beta) = T_{2} \\
        \;\ldots & \\
      (\frac{P - 1}{K} + \gamma(K+1) \cdot n_{s_M} - 2) \cdot (\alpha + \frac{m_{M}}{n_{s_2}} \cdot \beta) + (P - 1) \cdot (\alpha + \frac{m_{M}}{n_{s_{M}}} \cdot \beta) = T_{M} \\
   \end{cases}\\
   \MoveEqLeft[0]\text{}
  \\
   & 
    \Downarrow 
    \\ 
       & \begin{cases}
        \alpha + \frac{m_{1}}{n_{s_1}} \cdot \beta = \frac{T_{1}}{\frac{(P-1)\cdot(1+K)}{K} + \gamma(K+1) \cdot n_{s_{1}} - 2}\\
        \alpha + \frac{m_{2}}{n_{s_2}} \cdot \beta = \frac{T_{2}}{\frac{(P-1)\cdot(1+K)}{K} + \gamma(K+1) \cdot n_{s_{2}} - 2}\\	
        \;\ldots & \\
        \alpha + \frac{m_{M}}{n_{s_M}} \cdot \beta = \frac{T_{M}}{\frac{(P-1)\cdot(1+K)}{K} + \gamma(K+1) \cdot n_{s_{M}} - 2}\\	
     \end{cases} \\
     \MoveEqLeft[0]\text{}
  \end{flalign*}
    \caption{A system of $M$ non-linear equations with $\alpha$, $\beta$, and $\gamma(K+1)$ as unknowns, derived from $M$ communication experiments, each consisting of the execution of the  K-Chain tree broadcast algorithm, broadcasting a message of size $m_i$ ($i=1,...,M$) from the root to the remaining $P-1$ processes, followed by the linear gather algorithm without synchronisation, gathering messages of size $m_s$ (segment size) on the root. The execution times, $T_i$, of these experiments are measured on the root. \\
Given $\gamma(K+1)$ is evaluated  separately, the system becomes a system of $M$ linear equations with $\alpha$ and $\beta$ as unknowns.}
    \label{fig:systemoflinearequations} 
\end{figure*}

\section{Estimation of model parameters}
\label{sec:designofcommexperiments}

\subsection{Introduction to the estimation method}

In the most general case, the analytical model of an Open MPI collective algorithm uses three platform parameters -- $\alpha$, $\beta$, and $\gamma (p)$. The traditional state-of-the-art approach to estimation of $\alpha$ and $\beta$ would be to find these parameters from a number of point-to-point communication experiments. Namely, the time of a round-trip of a message of size $m$, $RTT(m)$, is measured for a wide range of $m$. From these experiments, a system of linear equations with $\alpha$ and $\beta$ as unknowns is derived. Then, linear regression is applied to find $\alpha$ and $\beta$. The found values of $\alpha$ and $\beta$ would be then used in all analytical predictive formulas. 

This approach yields a unique single pair of $(\alpha, \beta)$ for each target platform. Unfortunately, with $\alpha$ and $\beta$ found this way, not all our analytical formulas will be accurate enough to be used for accurate selection of the best performing collective algorithm. Using non-linear regression does not improve the situation as function $RTT(m)$ is typically near linear. Therefore, we propose to estimate the model parameters separately for each collective algorithm. More specifically, we propose to design a specific communication experiment for each collective algorithm, so that the algorithm itself would be involved in the execution of the experiment. Moreover, the execution time of this experiment must be dominated by the execution time of this collective algorithm. Then, we conduct a number of experiments on the target platform for a range of numbers of processors, $p$, and message sizes, $m$. From those experiments, we can derive a sufficiently large number of equations with $\alpha$, $\beta$, and $\gamma(p)$ as unknowns, and then use an appropriate solver to find their values. 

Unfortunately, when applied straightforwardly, this approach yields a system of \textit{non-linear} equations like the one shown in Figure \ref{fig:systemoflinearequations}. This non-linearity makes the task of estimation of the parameters mathematically very difficult, because we need to solve a large system of non-linear equations. 

Our approach to this problem is the following. As the non-linearity is caused by multiplicative terms involving $\gamma (p)$, we separate the estimation of $\gamma(p)$ from the estimation of $\alpha$ and $\beta$. Namely, we assume that $\gamma(p)$ is algorithm-independent
and design a separate communication experiment for its estimation.  The  values of $\gamma(p)$ found from this experiment are then used as known coefficients in the algorithm-specific systems of equations for $\alpha$ and $\beta$. 
We present this approach  in Sections \ref{subsec:mesurementofgamma} and \ref{subsec:measurementofalphabeta}.

\subsection{Estimation of $\gamma(p)$}
\label{subsec:mesurementofgamma}

The model parameter $\gamma(p)$ appears in the formula estimating the execution time of the  linear tree broadcast algorithm with non-blocking communication, which is only used for broadcasting of a segment in the tree-based segmented broadcast algorithms. Thus, in the context of Open MPI, the  linear tree broadcast algorithm with non-blocking communication will always broadcast a message of size $m_s$ to a relatively small number of processes.

According to Formula \ref{equ:gammafunction},

\begin{center}
  $\gamma (p) = \frac{T^{nonblock}_{linear\_bcast}(p, m_s)}{T_{p2p}(m_s)} =  \frac{T^{nonblock}_{linear\_bcast}(p, m_s)}{T^{nonblock}_{linear\_bcast}(2, m_s)}$ .
\end{center}

\noindent Therefore, in order to estimate $\gamma(p)$ for a given range of the number of processes, $p \in \{2,...,P\}$, we need to a method for estimation of $T^{nonblock}_{linear\_bcast}(p, m_s)$. We use the following method:

\begin{itemize}
  \item For each $2 \leq p \leq P$, we measure on the root the execution time $T_1(p, N)$ of $N$ successive calls to the  \textit{linear tree  with non-blocking communication} broadcast routine separated by barriers. The routine broadcasts a message of size $m_s$.
  \item We estimate $T^{nonblock}_{linear\_bcast}(p, m_s)$ as $T_2(p) = \frac{T_1(p, N)}{N}$.
\end{itemize} 


The experimentally obtained discrete function $\frac{T_2(p)}{T_2(2)}$ is used  as a platform-specific but algorithm-independent  estimation of $\gamma(p)$.

From our experiments, we observed that the discrete estimation of $\gamma(p)$ is near linear. Therefore, as an alternative for platforms with very large numbers of processors, we can build by linear regression a linear approximation of the discrete function $\frac{T_2(p)}{T_2(2)}$, obtained for a representative subset of the full range of $p$, and use this linear approximation as an analytical estimation of $\gamma(p)$.

\subsection{Estimation of algorithm specific $\alpha$ and $\beta$}
\label{subsec:measurementofalphabeta}
To estimate the model parameters $\alpha$ and $\beta$ for a given collective algorithm, we design a communication experiment, which starts and finishes on the root (in order to accurately measure its execution time using the root clock), and involves the execution of the modelled collective algorithm so that the total time of the experiment would be dominated by the time of its execution. 

For example, for all broadcast algorithms, the communication experiment consists of a broadcast of a message of size $m$ (where $m$ is a multiple of segment size $m_s$), using the modelled broadcast algorithm, followed by a \textit{linear-without-synchronisation} gather algorithm, gathering messages of size $m_s$ on the root. The execution time of this experiment on $p$ nodes,  $T_{bcast\_exp}(p,m)$, can be estimated as follows:

\begin{equation}
  T_{bcast\_exp}(p,m) = T_{bcast\_alg}(p,m) + T_{linear\_gather}(p, m_s) 
  \label{equ:experiemntsanalyticalformula}
\end{equation}

Using analytical formulas from Section \ref{sec:implementationdrivenmodel} for $T_{bcast\_alg}(p,m)$ and $T_{linear\_gather}(p, m_s)$, for each combination of $p$ and $m$ this experiment will yield one linear equation with $\alpha$ and $\beta$ as unknowns. By repeating this experiment with different $p$ and $m$, we obtain a system of linear equations for $\alpha$ and $\beta$. Each equation in this system can be represented in the canonical form,  $\alpha + \beta \times m_i = T_i$ ($i=1,...,M$).  Finally, we use the  least-square regression to find $\alpha$ and $\beta$, giving us the best linear approximation $\alpha + \beta \times m$ of the discrete function $f(m_i) = T_i$  ($i=1,...,M$).   


 Figure \ref{fig:systemoflinearequations} shows a system of linear equations built for the K-Chain tree broadcast algorithm for our experimental platform. To build this system, we used the same $P$ nodes in all experiments but varied the message size $m \in \{m_1, ..., m_M\}$. With $M$ different message sizes, we obtained a system of $M$ equations.  The number of nodes, $P$, was approximately equal to the half of the total number of nodes. We observed that the use of larger numbers of nodes in the experiments will not change the estimation of $\alpha$ and $\beta$.  



\begin{figure*}
  \centering
  \begin{subfigure}[t]{0.5\textwidth}
  \includegraphics[width=1\linewidth]{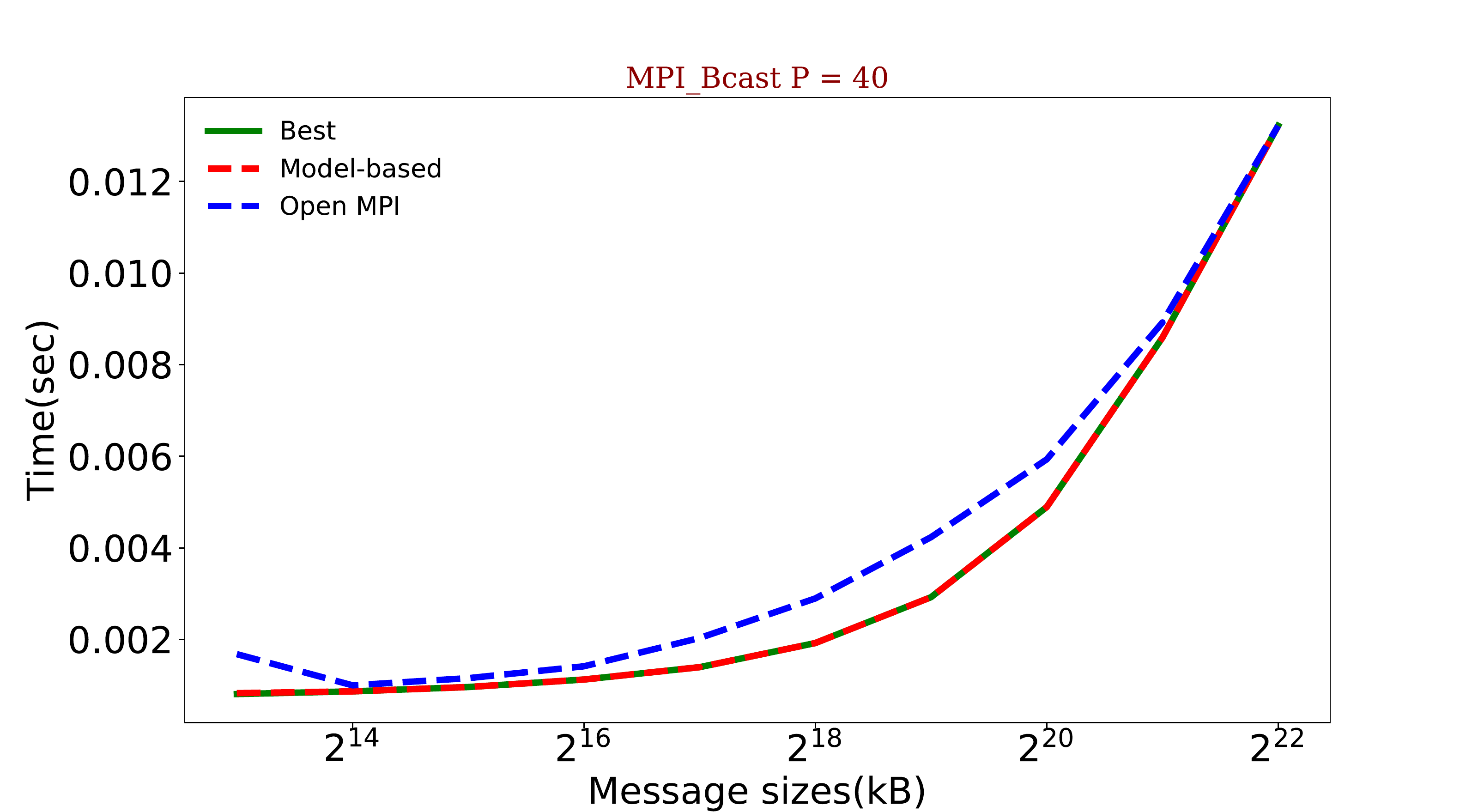}
  \caption{}
  \label{subfig:bcast_40_sel}
  \end{subfigure}%
  ~
  \begin{subfigure}[t]{0.5\textwidth}
  \includegraphics[width=1\linewidth]{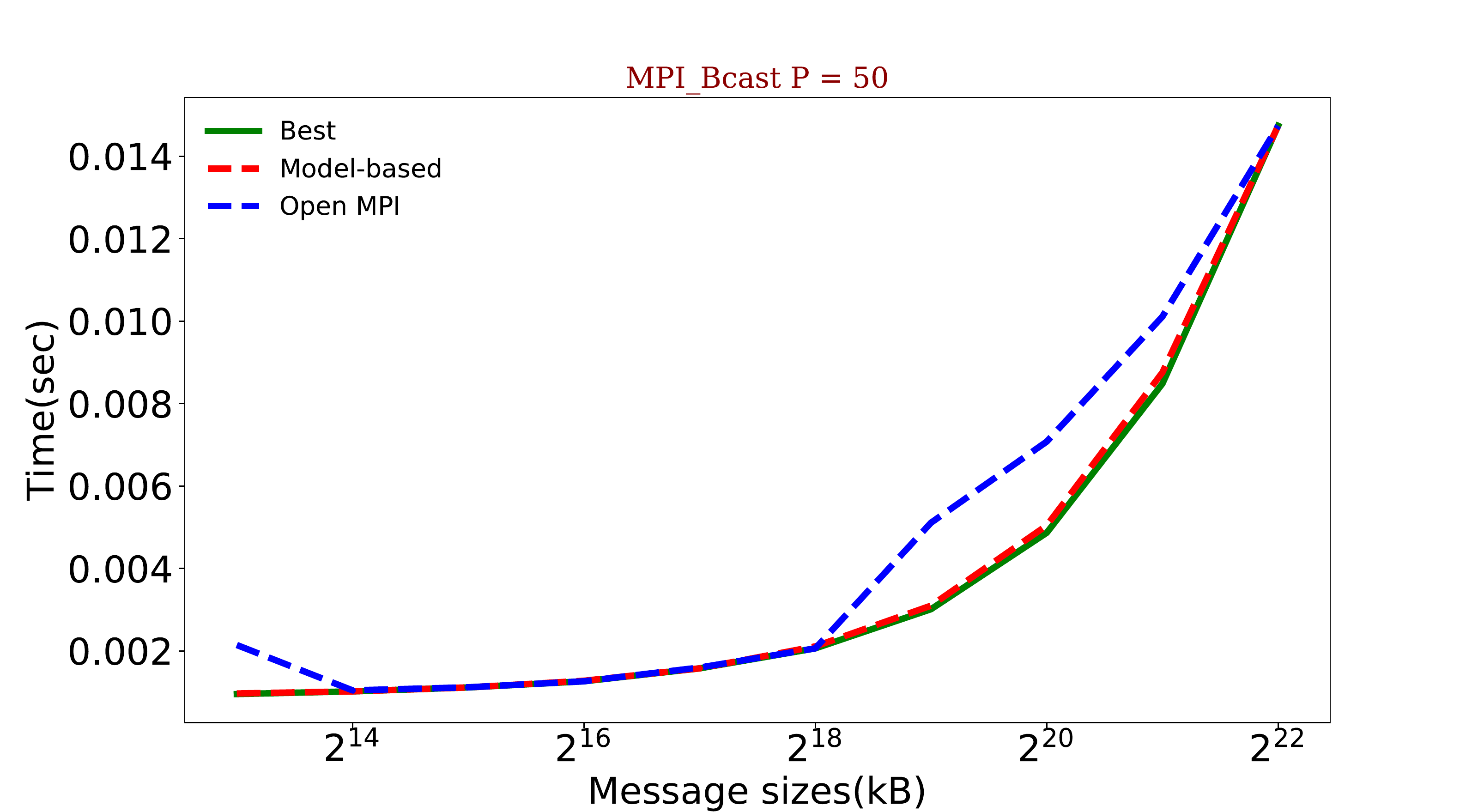}
  \caption{}
  \label{subfig:bcast_50_sel}
  \end{subfigure}
  
  \begin{subfigure}[t]{0.5\textwidth}
  \includegraphics[width=1\linewidth]{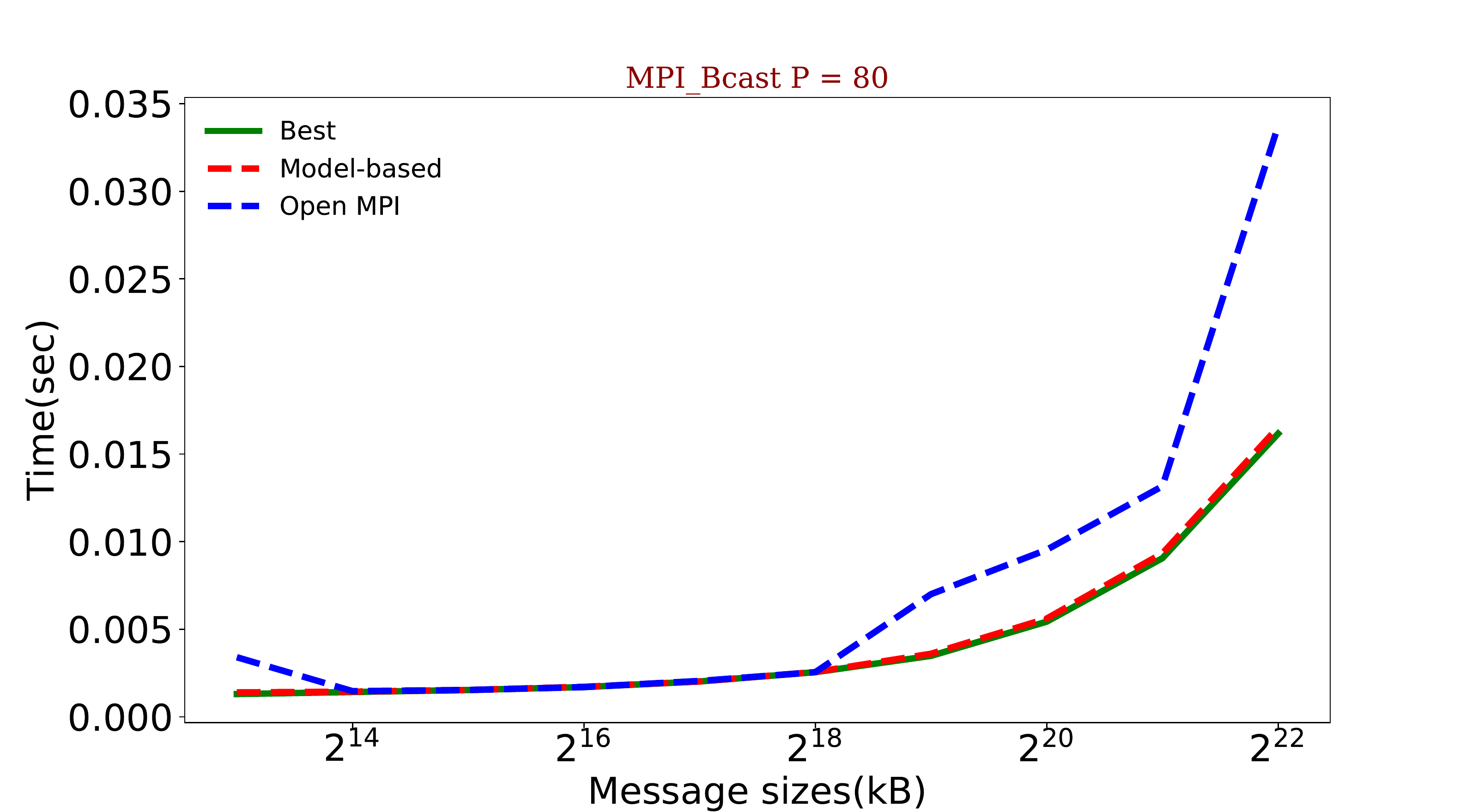}
  \caption{}
  \label{subfig:bcast_80_sel}
  \end{subfigure}%
  ~
  \begin{subfigure}[t]{0.5\textwidth}
  \includegraphics[width=1\linewidth]{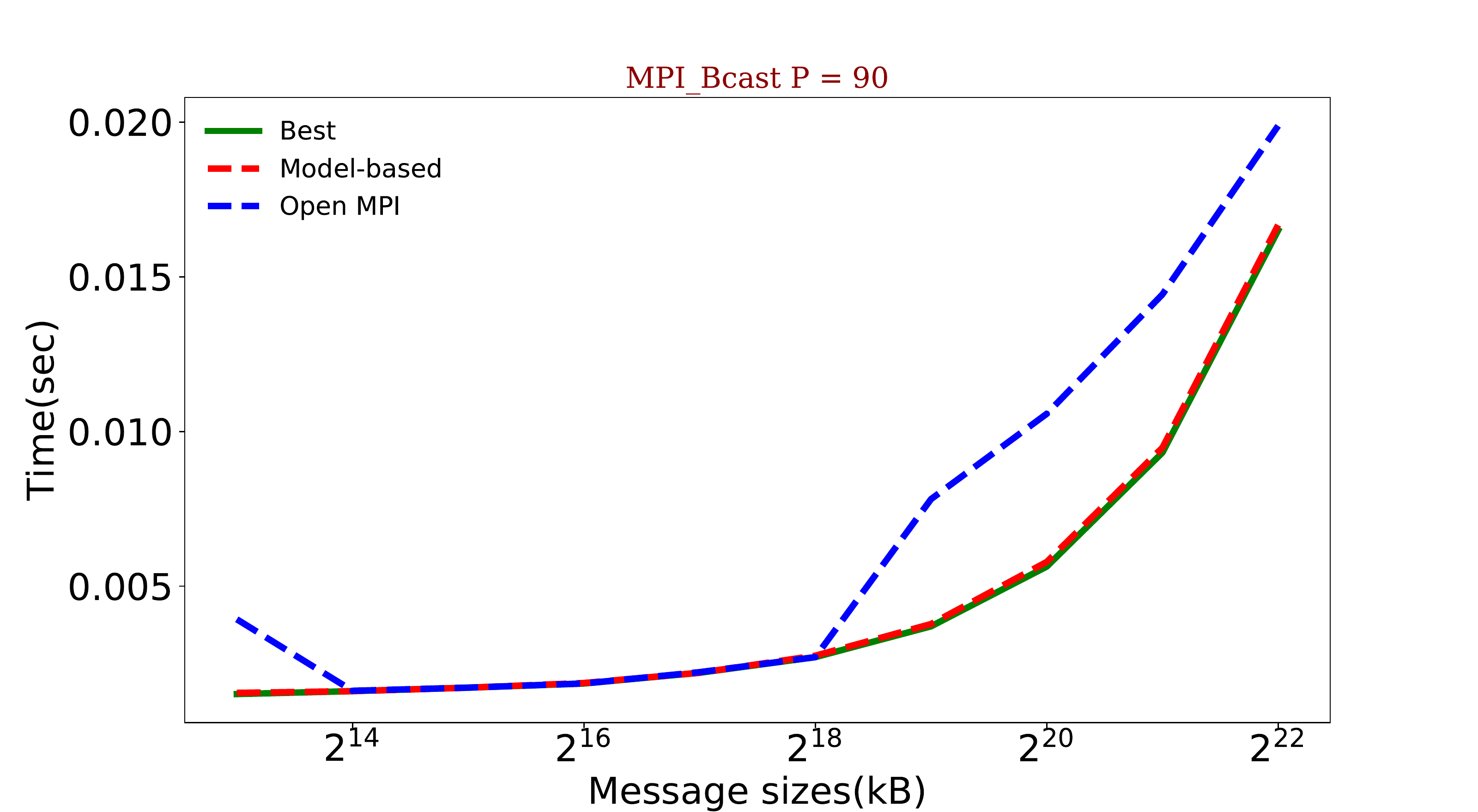}
  \caption{}
  \label{subfig:bcast_90_sel}
  \end{subfigure}
  
  \begin{subfigure}[t]{0.5\textwidth}
  \includegraphics[width=1\linewidth]{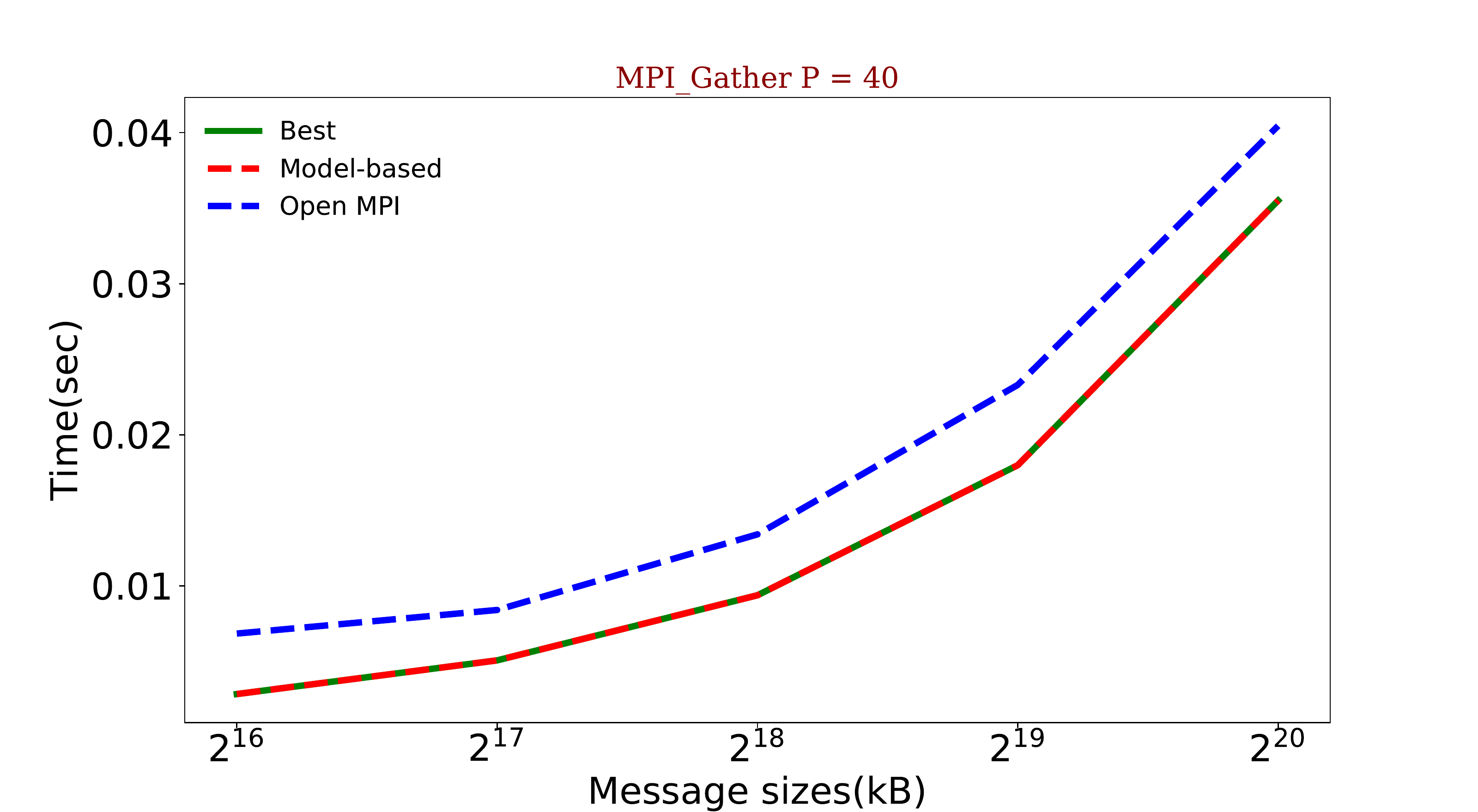}%
  \caption{}
  \label{subfig:gather_40_sel}
  \end{subfigure}%
  ~
  \begin{subfigure}[t]{0.5\textwidth}
  \includegraphics[width=1\linewidth]{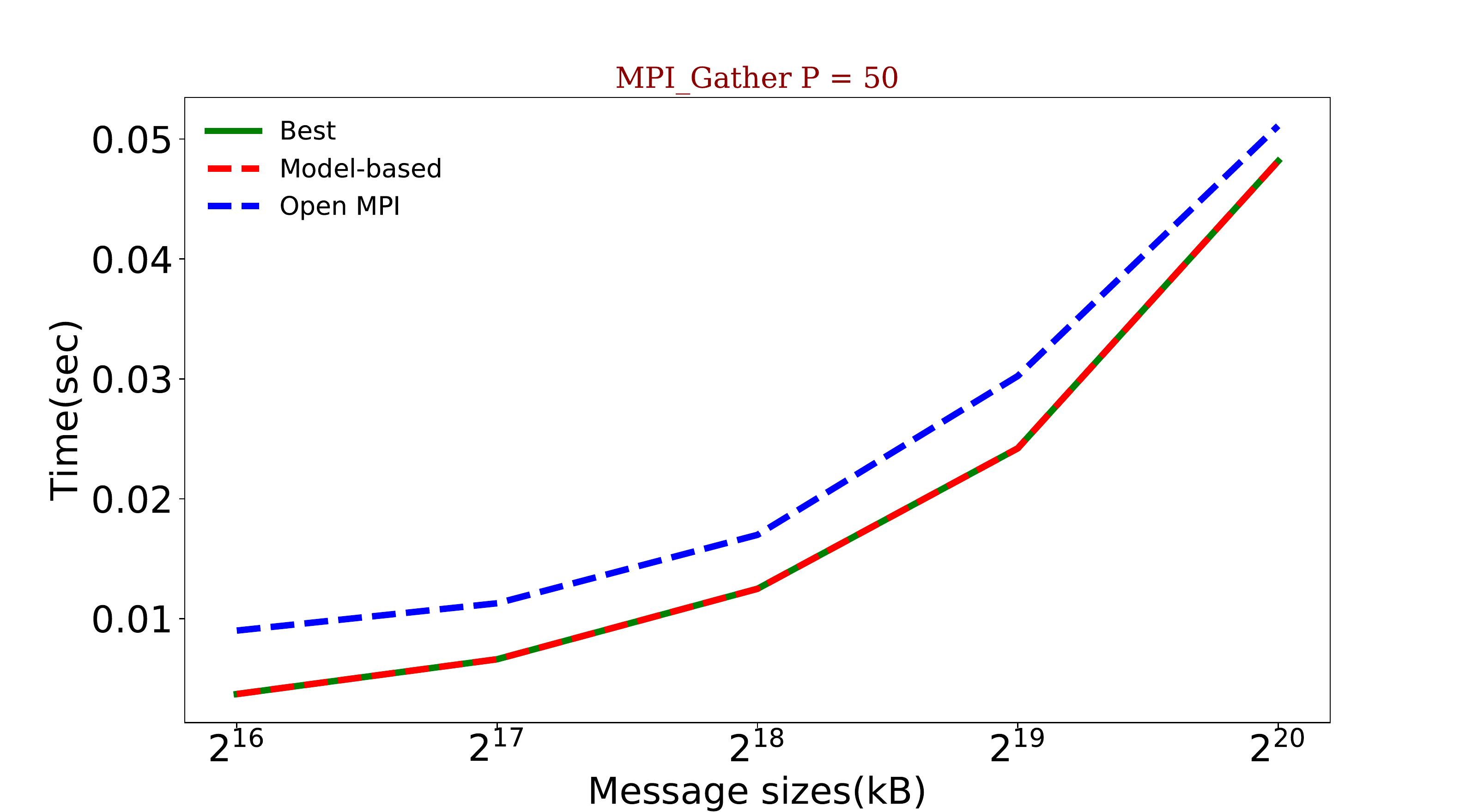}%
  \caption{}
  \label{subfig:gather_50_sel}
  \end{subfigure}
  \begin{subfigure}[t]{0.5\textwidth}
  \includegraphics[width=1\linewidth]{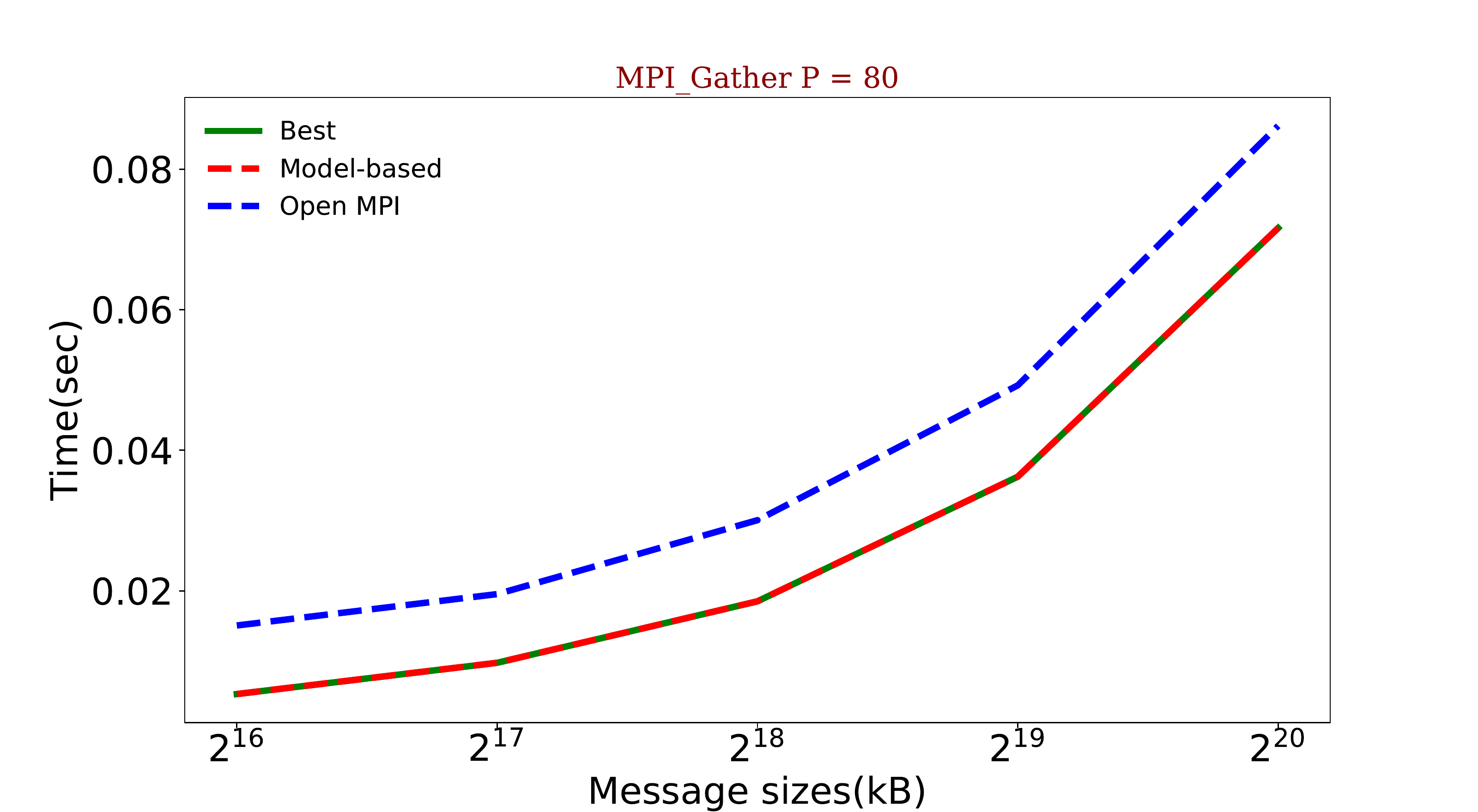}
  \caption{}
  \label{subfig:gather_80_sel}
  \end{subfigure}%
  ~
  \begin{subfigure}[t]{0.5\textwidth}
  \includegraphics[width=1\linewidth]{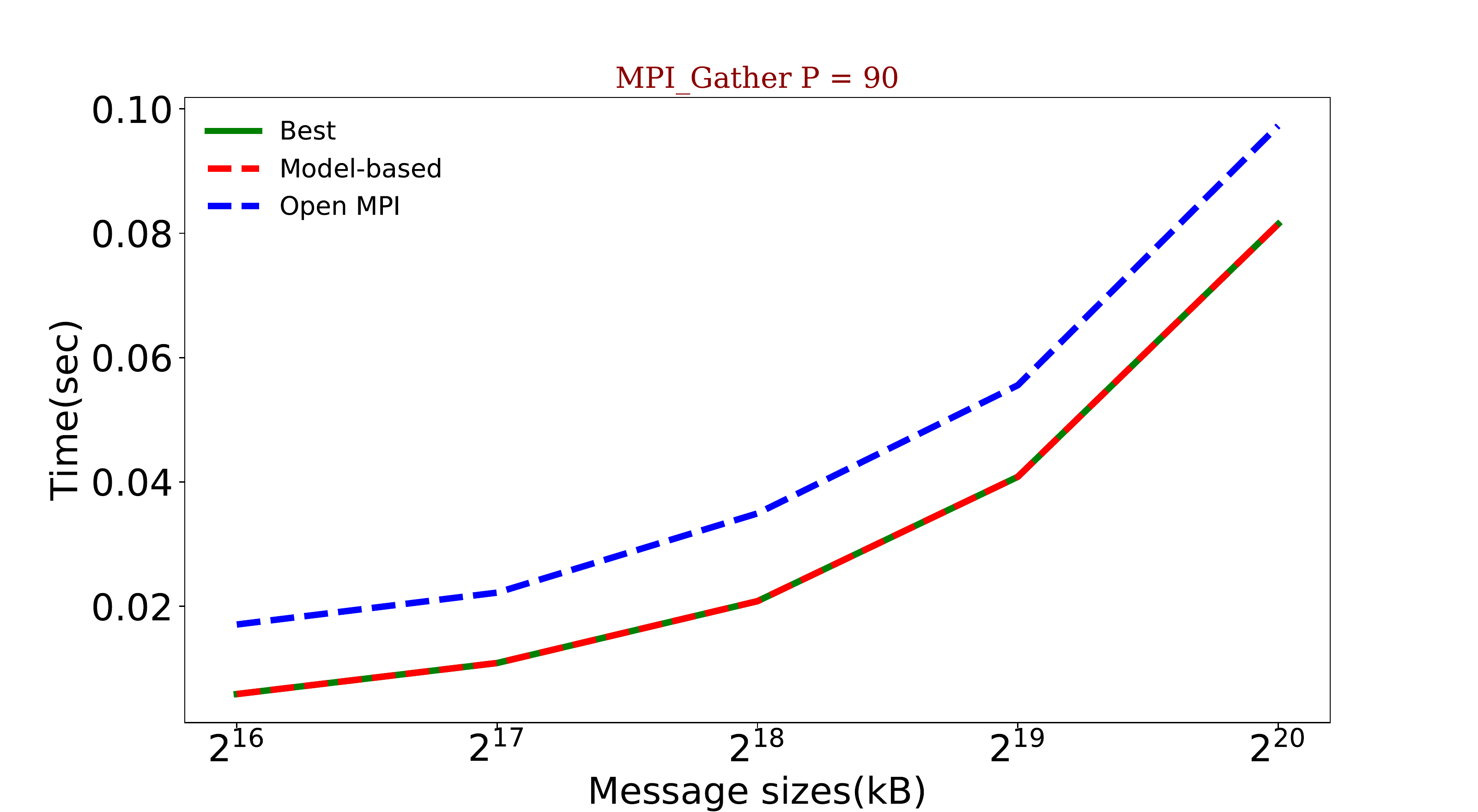}
  \caption{}
  \label{subfig:gather_90_sel}
  \end{subfigure}
  \caption{Comparison of the selection accuracy of the Open MPI decision function and the proposed model-based method for MPI\_Bcast  (\ref{subfig:bcast_40_sel} - \ref{subfig:bcast_90_sel}) and MPI\_Gather (\ref{subfig:gather_40_sel} - \ref{subfig:gather_90_sel}).}
\label{fig:algselectiongather}
\end{figure*}

\section{Experimental results and analysis}
\label{sec:experimentalresults}

This section presents experimental evaluation of the proposed approach to selection of optimal collective algorithms using Open MPI broadcast and gather operations as an example. 


\subsection{Experiment setup}
For experiments, we use Open MPI 3.1.3 running on a dedicated Grisou cluster of the Nancy site of the Grid`5000 infrastructure \cite{grid5000}. The  cluster consists of 51 nodes each with 2 Intel Xeon E5-2630 v3 CPUs (8 cores/CPU), 128GB RAM, 2x558GB HDD, interconnected via 10Gbps Ethernet. 

To make sure that the experimental results are reliable, we follow a detailed methodology: 1) We make sure that the cluster is fully reserved and dedicated to our experiments. 2) For each data point in the execution time of collective algorithms, the sample mean is used, which is calculated by executing the application repeatedly until the sample mean lies in the 95\% confidence interval and a precision of 0.025 (2.5\%) has been achieved. We also check that the individual observations are independent and their population follows the normal distribution.  For this purpose, MPIBlib \cite{lastovetsky2008mpiblib} is used.

\begin{table}[b]
\centering
\begin{tabular}{ | c | c | } 
\hline
Number of processes (p) & $\gamma(p)$ \\ 
\hline
3 & 1.114 \\ 
\hline
4 & 1.219 \\ 
\hline
5 & 1.283 \\ 
\hline
6 & 1.451 \\ 
\hline
7 & 1.540 \\ 

\hline
\end{tabular}
\caption{Estimated values of $\gamma(p)$ on Grisou cluster.}
\label{tab:gammavalues}
\end{table}





\begin{table}[b]
\centering
\begin{tabular}{ | p{4cm} | c | c | } 
\hline
Collective algorithm & $\alpha (sec)$ &$\beta$ ($\frac{sec}{byte}$) \\ 
\hline
\multicolumn{3}{|c|}{\textbf{Broadcast}} \\
\hline
Linear tree & \num{2.2e-12} & \num{1.8e-08} \\ 
\hline
K-Chain tree & \num{5.7e-13} &\num{4.7e-09}\\ 
\hline
Chain tree& \num{6.1e-13} & \num{4.9e-09} \\ 
\hline
Split-binary tree & \num{3.7e-13} & \num{3.6e-09} \\ 
\hline
Binary tree & \num{5.8e-13} & \num{4.7e-09} \\ 
\hline
Binomial tree & \num{5.8e-13} & \num{4.8e-09} \\ 
\hline
\multicolumn{3}{|c|}{\textbf{Gather}} \\
\hline
Linear tree without synchronisation & \num{1.5e-05} & \num{1.5e-09} \\ 
\hline
Binomial tree & \num{1.2e-04} & \num{8.6e-10} \\ 
\hline
Linear tree with synchronisation & \num{5.9e-05} & \num{9.4e-10} \\ 
\hline
\end{tabular}
\caption{Estimated values of $\alpha$ and $\beta$ for the Grisou cluster and Open MPI  broadcast and gather algorithms.}
\label{tab:alphabetavalues}
\end{table}


In our communication experiments, MPI programs use the one-process-per-CPU configuration, and the maximal total  number of processes is equal to 90. The message segment size, $m_s$, for segmented broadcast algorithms is set to 8KB. This segment size is commonly used for segmented broadcast algorithms in Open MPI. Selection of optimal segment size is out of the scope of this paper.


   

\subsection{Experimental estimation of model parameters}
\label{subsec:expmeasurementgamma}

Estimation of  parameter $\gamma(p)$ for our experimental platform follows the method presented in Section \ref{subsec:mesurementofgamma}. With the maximal number of processes equal to 90, the maximal
number of children in the linear tree broadcast algorithm with non-blocking communication,  used in the segmented Open MPI broadcast algorithms, will be equal to  seven. Therefore, the number of processes in our communication experiments ranges from 2 to 7. By definition, $\gamma(2) = 1$. The estimated values of $\gamma(p)$ for $p$ from 3 to 7 are given in Table \ref{tab:gammavalues}.



After estimation of $\gamma(p)$, we conduct communication experiments to estimate algorithm-specific values of parameters $\alpha$ and $\beta$ for six broadcast algorithms and three gather algorithms following the method described in Section \ref{subsec:measurementofalphabeta}. In all the experiments we use the same number of processes, 40. The message size , $m$, varies in the range from  8KB to 4MB in the broadcast experiments, and from 64KB to 1MB in the gather experiments. We use $10$ different sizes for broadcast algorithms, $\{m_i\}^{10}_{i=1}$, and $5$ different sizes for gather algorithms, $\{m_i\}^{5}_{i=1}$, separated by a constant step in the logarithmic scale, $\log{m_{i-1}} - \log{m_i} = const$. Thus, for each collective algorithm, we obtain a system of 10 linear equations with $\alpha$ and $\beta$ as unknowns. We use the Huber regressor \cite{huber1992robust} to find their values from the system.

The values of parameters $\alpha$ and $\beta$ obtained this way can be found in Table \ref{tab:alphabetavalues}. We can see that the values  of  $\alpha$ and $\beta$ do vary depending on the collective algorithm, and the difference is more significant between algorithms implementaing different collective operations. The results support our original hypothesis that the average execution time of a point-to-point communication will very much depend on the context of the use of the point-to-point communications in the algorithm. One interesting example is the Split-binary tree and Binary tree broadcast algorithms. They both use the same virtual topology, but the estimated time of a point-to-point communication, $\alpha + \beta \times m$, is smaller in the context of the Split-binary one. This can be explained by a higher level of parallelism of the Split-binary algorithm, where a significant part of point-to-point communications is performed in parallel by a large number of independent pairs of processes from the left and right subtrees. 

\subsection{Accuracy of selection of optimal collective algorithms using the constructed analytical performance models}

The constructed analytical performance models  of the Open MPI broadcast and gather collective algorithms are designed for the use in the MPI\_Bcast and  MPI\_Gather routines for efficient and accurate runtime selection of the optimal  algorithm, depending on the number of processes and the message size. While the efficiency is evident from the low complexity of the analytical formulas derived in Section \ref{sec:implementationdrivenmodel}, the experimental results on the accuracy are presented in this section.

Figure \ref{fig:algselectiongather} shows the results of our experiments for MPI\_Bcast and  MPI\_Gather. For both operations, we present results of experiments with four different numbers of processes ranging from 40 to 90.  The message size, $m$, varies in the range from  8KB to 4MB in the broadcast experiments, and from 64KB to 1MB in the gather experiments. We use $10$ different sizes for broadcast algorithms, $\{m_i\}^{10}_{i=1}$, and $5$ different sizes for gather algorithms, $\{m_i\}^{5}_{i=1}$, separated by a constant step in the logarithmic scale, $\log{m_{i-1}} - \log{m_i} = const$. The graphs show the execution time of the collective operation as a function of the message size. Each data point on a blue line shows the performance of the algorithm selected by the Open MPI decision function for the given operation, number of processes and message size. Each point on a red line shows the performance of the algorithm selected by our decision function, which uses the constructed analytical models. Each point on a green line shows the performance of the best Open MPI algorithm  for the given collective operation, number of processes and message size.

As can be seen from the results, the Open MPI selection is mostly inaccurate for  MPI\_Bcast and never accurate for MPI\_Gather. It is also evident from the graphs that this inaccuracy can be very costly in terms of performance. On the other hand, while equally efficient with the Open MPI selection method,  our selection method turned out to be 100\% accurate, always selecting the best algorithm. 



\section{Conclusions}
\label{sec:conclusions}

In this paper, we proposed a novel model-based approach to automatic selection of optimal algorithms for MPI collective operations, which proved to be both efficient and accurate. The novelty of the approach is two-fold. First, we proposed to derive  analytical models of collective algorithms from the code of their implementation rather than from high-level mathematical definitions. Second, we proposed to estimate model parameters separately for each algorithm, using a communication experiment, where the execution of the algorithm itself dominates the execution time of the experiment. 

We also developed this approach into a detailed method and applied it to Open MPI 3.1 and its MPI\_Bcast and MPI\_Gather operations. We experimentally validated this method on a cluster of 51 dual-processor nodes and demonstrated its accuracy and efficiency. These results suggest that the proposed approach, based on analytical performance modelling of collective algorithms,  can be successful in the solution of the problem of accurate and efficient runtime selection of optimal algorithms for MPI collective operations.

\ifCLASSOPTIONcompsoc
  \section*{Acknowledgments}
\else
  \section*{Acknowledgment}
\fi
This publication has emanated from research conducted with the financial support of Science Foundation Ireland (SFI) under Grant Number 14/IA/2474. 

Experiments presented in this paper were carried out using the Grid'5000 experimental testbed, being developed under the INRIA ALADDIN development action with support from CNRS, RENATER and several Universities as well as other funding bodies (see \href{https://www.grid5000.fr}{https://www.grid5000.fr}).

\ifCLASSOPTIONcaptionsoff
  \newpage
\fi

\bibliographystyle{IEEEtran}
\bibliography{IEEEabrv,collalgselection}

\end{document}